\documentstyle[apj,times,psfig]{article}

\def\simg{\mathrel{\rlap{\raise 0.511ex \hbox{$>$}}{\lower 0.511ex \hbox{$\sim$}}}}
\def\siml{\mathrel{\rlap{\raise 0.511ex \hbox{$<$}}{\lower 0.511ex \hbox{$\sim$}}}}
\def\deg{^{\rm o}} \def\h1{\hspace*{-1mm}} \def\p2{\pi/2} \def\cm2{{\rm cm}^{-2}}

\begin{document}

\parskip 4pt
\topmargin 1cm

\vspace*{-10mm}
\title{The 7$\frac{2}{3}$-Year Collection of Well-Monitored Fermi-LAT GRB Afterglows}

\author{A. Panaitescu}
\affil{ Space \& Remote Sensing, MS B244, Los Alamos National Laboratory, Los Alamos, NM 87545, USA}

\vspace*{2mm}
\begin{abstract}
  We present the light-curves and spectra of 24 afterglows that have been monitored by Fermi-LAT at 0.1--100 GeV
  over more than a decade in time. All light-curves (except 130427) are consistent with a single power-law starting 
  from their peaks, which occurred, in most cases, before the burst end.
  The light-curves display a brightness-decay rate correlation, with all but one (130427) of the bright afterglows 
  decaying faster than the dimmer afterglows. We attribute this dichotomy to a quick deposition of the relativistic 
  ejecta energy in the external-shock for the brighter/faster-decaying afterglows and to an extended energy-injection 
  in the afterglow shock for the dimmer/slower-decaying light-curves.
  The spectra of six afterglows (090328, 100414, 110721, 110731, 130427, 140619B) indicate the existence of a harder 
  component above a spectral dip/ankle at energy 0.3--3 GeV, offering evidence for an inverse-Compton emission at 
  higher energies, and suggesting that the harder power-law spectra of five other LAT afterglows (130327B, 131231, 
  150523, 150627, 160509) could also be inverse-Compton, while the remaining softer LAT afterglows should be synchrotron.
  Marginal evidence for a spectral break and softening at higher energies is found for two afterglows 
 (090902B and 090926).
\end{abstract}

\keywords{radiation mechanisms: non-thermal, relativistic processes, shock waves -- gamma-ray burst: general, individual}

\section{Introduction}

 Using the Fermi Science tools and the Fermi-LAT photon database, we identify 24 afterglows that have been followed
by LAT over more than a decade in time. We construct the light-curves and spectra of those afterglows, visually search
for correlations, fit them with simple power-laws (and broken power-laws for spectra), and compare the resulting
fit indices/exponents with the expectations for the forward-shock model, all in an broad-brush attempt to identify
generic features of that model that could explain the light-curve correlations and spectra features.

 Our selection of well-monitored LAT afterglows, meaning afterglows monitored for more than a decade in time 
implies the exclusion of those afterglows that had a too fast fall-off after their first detection, especially 
if such afterglows were dim at early times (during the prompt-emission phase). From the First Fermi-LAT GRB catalog 
(Ackermann et al 2013a), 10 afterglows that display a long-monitored light-curve are included in 
the current sample. To those, we add another 14 afterglows with well-monitored light-curves, the current sample
containing 21 percent of the 108 afterglows detected by LAT during the first 7$\twothirds$ years. In other words, LAT 
detects on average one GRB afterglow per month, but only one out of four (about 3/year on average) is a 
well-monitored afterglow. 

 In addition to the First LAT GRB catalog, the LAT light-curves and spectra of some of the afterglows in our 
sample have been previously published in other articles: 080916C (Abdo et al 2009a), 090217 (Ackermann et al
2010b), 090510 (Ackermann et al 2010a, Ghirlanda, Ghisellini, Nava 2010), 090902B (Abdo et al 2009b), 090926 
(Swenson et al 2010, Ackermann et al 2011), 110721 (Axelsson et al 2012), 110731 (Ackermann et al 2013b), 
130427 (Fan et al 2013, Tam et al 2013, Ackermann et al 2014), 130821 (Liang et al 2014), 131231 (Liu et al 2014). 
The afterglows 090510 and 131108 have also been observed above 100 MeV by AGILE (Giuliani et al 2010, Giuliani et al 2014).

\section{How Afterglow Light-Curves and Spectra are Calculated and Fit}

 The "event" and "spacecraft" data used here were downloaded from the Fermi-LAT database at 
{\sl http://fermi.gsfc.nasa.gov/cgi-bin/ssc/LAT/LATDataQuery.cgi} and processed using the Fermi Science
tools at {\sl fermi.gsfc.nasa.gov/ssc/data/analysis/scitools/references.html}. The event files
are downloaded with "extended" option that includes "transient"-class photons, having more background. 
The transient-class photons are used in the construction of afterglow light-curves, while spectra are 
obtained using "source"-class photons that have a lower background.

 Selection cuts on the event data files are done using the {\sc gtselect} tool. The cuts include:
selection of photon-class (16 for transient, 128 for source) above 100 MeV, within a Search Region radius 
of $10\deg$  around the burst location, starting at the burst trigger time, and exclusion of photons close 
to Earth's limb (at zenith angle $112\deg$). Events within the time range when the data are valid are
retained using the {\sc gtmktime} tool. Tool {\sc gtbindef} is used to create the logarithmic time and 
energy-grids for light-curves and spectra. Photon counts with specified time-binning are obtained using 
the {\sc gtbin[lc]} tool and the full-exposure (LAT effective-area, i.e. integrated over the afterglow spectrum
and corresponding to the photon incidence angle at each time, multiplied by the exposure-time corresponding 
to each time-bin) are added to the light-curve file using the {\sc gtexposure} tool and the appropriate
instrument response function: {\sc p8r2\_transient020\_v6} for transient-class photons and
{\sc p8r2\_source\_v6} for source-class photons. Photon fluxes above 100 MeV are calculated by dividing 
photon counts by the full-exposure of each time-bin. Photon counts on a given time and energy-grid are
calculated with {\sc gtbin[pha2]} tool, from where spectra are obtained by dividing by the full-exposure 
calculated with the {\sc gtexposure} tool on the spectral time-binning. Spectra are corrected for the
LAT effective-area dependence on the photon energy (fig 15 of Atwood et al 2009), which for photons at 
normal incidence can be well-approximated by $A(E<1GeV)= A_0 [1+0.6 \log (E/GeV)]$ and $A(E>1GeV) \simeq 
A_0$, the coefficient 0.6 having a weak dependence on the incidence angle $\theta$ (e.g. it is 0.7 for
$\theta = 60\deg$).
 
  Figure 1 shows the light-curves and spectra of GRB afterglow 130427, and assesses the effects of using
transient or source-class photons, as well as that of the size of the region around the source over which
photons are counted. Detector background photons account for 46 percent of the transient-class photons 
in the lowest energy channel (100 MeV) shown in the right-upper panel, with that proportion decreasing 
strongly with photon energy, such that there are no background photons above 2 GeV (i.e. all transient-class
photons are source-class). Consequently, spectra built with source-class photons are slightly harder 
than those constructed with transient-class photons. The effect on light-curves (left-upper panel) is 
often un-noticeable because, in the calculation of the full-exposure, we used the detector response-function 
corresponding to each photon class.

\begin{figure*}[t]
\centerline{\psfig{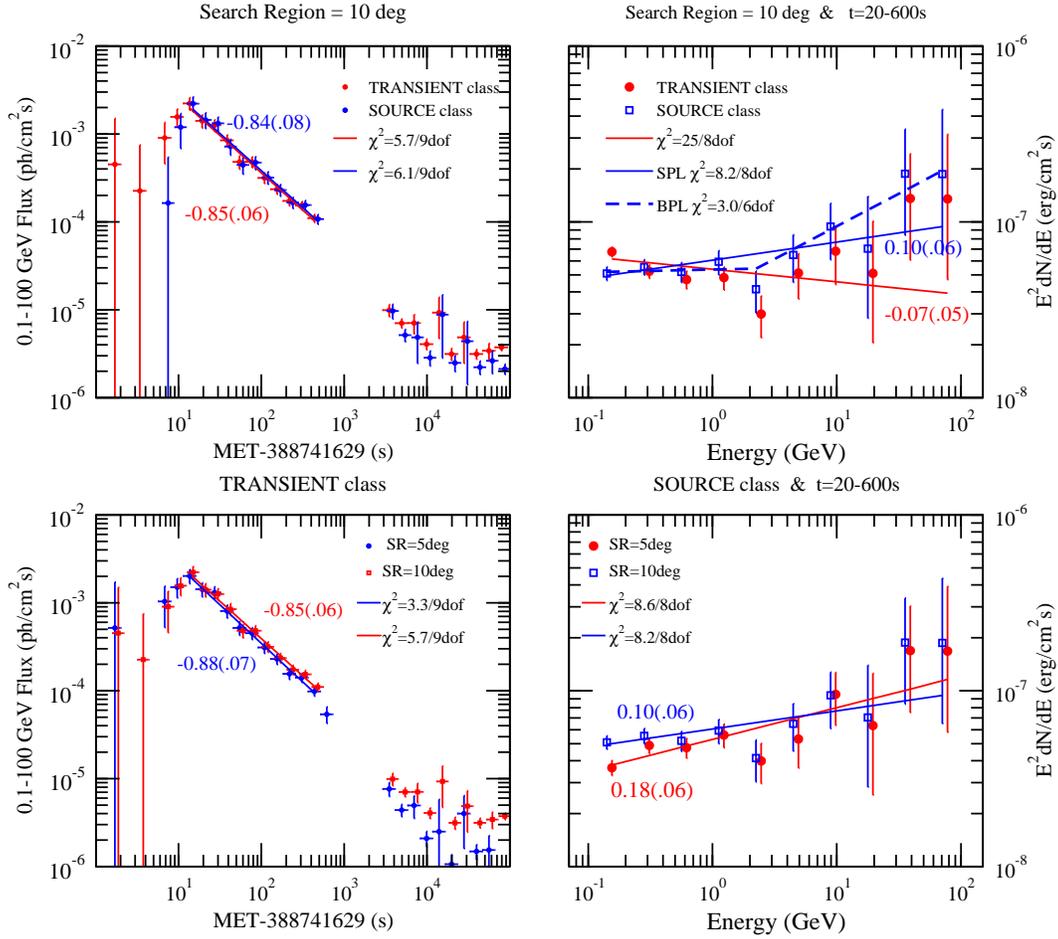}}
\figcaption{ Comparison of light-curves (left column) and spectra (right column) for the GRB afterglow 130427A 
       obtained with {\sc transient}-class and {\sc source}-class photons (upper row) and for a {\sc source region}
       of $5\deg$ and $10\deg$ (lower row). 
       {\sl Right row}: spectra $E^2(dN/dE) \propto E^{1-\beta}$ are fit with single power-laws, with the best-fit 
         exponent $1-\beta$ indicated. 
       {\sl Left row}: light-curves for photons above 100 MeV are fit with a single power-law $t^{-\alpha}$ 
       over the time-range 20--600s, the best-fit decay exponent $-\alpha$ and its uncertainty being indicated. 
       (At 600-3000s, the afterglow is behind by the Earth.) 
       After 10 ks, most LAT photons are the celestial background, consistent with the factor four difference 
       between the photon fluxes in source regions of $5\deg$ and $10\deg$ shown in the lower-left panel. 
       {\sl Upper panels}: transient-class photons contain more detector background than source-class photons, 
        but that does not change significantly the light-curve decay index $\alpha$ (upper-left panel). 
       Because most instrument background photons are at low energy, their removal leads to source-class spectra 
        that are harder than transient-class spectra by $\Delta \beta = -0.17 \pm 0.08$ (right-upper panel).
       {\sl Lower panels}: because $PSF(100MeV)=5\deg$, increasing the source-region size from $5\deg$ to $10\deg$ 
        increases slightly the photon flux but leaves unchanged the flux decay index $\alpha$ (lower-left panel); 
        adding that LAT's $PSF(E) \propto E^{-0.8}$, means that doubling the source-region size leads to a softer 
        spectrum, by $\Delta \beta = 0.16 \pm 0.11$ (lower-right panel).
       {\sl Upper-right panel}: the reduction of $\Delta \chi^2 = 2.8$ in the chi-square from a {\sl single} power-law 
        fit (blue solid line) to the 20--600 s source-class photon spectrum to a {\sl broken} power-law fit (dashed black line) 
        has a 1.8 percent probability of occurring by chance (according to the $F$-test). 
        The broken power-law best-fit has a break energy $E_* \simeq 2$ GeV and spectral slopes below and above
        the ankle are $\beta_{lo} = 1.0$ and $\beta_{hi} = 0.6$.
        Ackermann et al (2014) have identified a hard spectral component with $\beta = -1/3$ at earlier times,
        4.5--11.5 s, during the prompt emission phase.
  }
\end{figure*}

 Afterglow light-curves are calculated within a search area radius of $10\deg$ around the burst, 
but a search area of only $5\deg$ is used for dimmer afterglows that reach the background in less than 
a decade in time after the peak or after the first measurement. If that background is mostly the isotropic 
celestial background, then a reduction by a factor two in the search radius lowers the background by 
a factor four, which, for a flux decay $F \propto t^{-1}$, extends by 0.5 dex the measurement of the
afterglow power-law flux decay before is becomes lost in the background. An ever stronger reduction of
the background results if most of that background is from a steady source near the burst. That is the case 
for GRBs 110731, 130327B, 150902, and 150627, which are at several degrees from the Galactic plane, and 
for GRBs 150523 and 150627 which occurred at $7\deg$ off the Vela pulsar.

 The LAT point-spread function being $PSF(E) = 3.5\deg (E/100 MeV)^{-0.8}$ for on-axis photons, and
having a slow increase with the incidence angle (e.g. the coefficient is $5\deg$ for $\theta = 60\deg$ --
fig 17 of Atwood et al 2009), means that a source-region of $5\deg$ misses about 1/3 of the 100 MeV 
photons, while a $10\deg$ region loses about 5 percent of same photons, with the loss of photons being
negligible above 200 MeV, for which $PSF(E>200 MeV) < 2-3\deg$. Therefore, using a source region of
only $5\deg$ (for the reasons discussed above) leads to a small flux decrease (see light-curves in the
lower-right panel of Fig 1), and the corresponding loss of predominantly lower energy photons yields 
slightly harder spectra (as illustrated in the right-lower panel). 

 In summary of the above and of Figure 1, light-curves will be calculated using transient-class photons 
integrated over $5\deg$ or $10\deg$ around the source although neither the choice of photon class nor
the size of the source region affect much the index of the afterglow flux power-law decay, while spectra
are calculated using source-class photons integrated over $10\deg$ around the source, in order to obtain
a reliable spectral slope, although that aim will be too often defeated by the paucity of high-energy
photons (above 1 GeV), which will lead to uncertainties in the best-fit power-law exponent that exceed
the accuracy $\Delta \beta \siml 0.2$ allowed by the use of (lower background) source-class photons.


 The afterglow light-curves and spectra obtained with the Fermi Science tools as described above are
fit with power-laws, as is routinely done for GRB afterglows observed at lower photon energies: radio 
(1--100 GHz), optical (1--3 eV), or X-ray (1--10 keV). Power-law spectra are expected for the emission
from shock-accelerated electrons, whose distribution with energy is a power-law $dN_e/dE \propto E^{-p}$.
Then, synchrotron energy-spectra from such electrons will be a power-law $E (dN_\gamma/dE) \propto E^{-\beta}$ 
with the spectral index $\beta = (p-1)/2$. Adding to such power-law spectra a power-law 
dynamics of the source (i.e. Lorentz factor and number of emitting electrons that evolve as power-laws 
in observer-time) leads to power-law afterglow light-curves $dN_\gamma/dt \propto t^{-\alpha}$,
as predicted by M\'esz\'aros \& Rees (1997) for the synchrotron emission from an adiabatic forward-shock driven 
by a relativistic outflow in the circumburst medium (see also Kumar \& Barniol Duran 2009) and for the
synchrotron emission from cooling electrons that were accelerated by the reverse-shock crossing the
GRB outflow at early times. The power-law light-curves also stand for non-adiabatic shocks provided that
energy is added (Rees \& M\'esz\'aros 1998) to the external shock (reverse and forward shocks) in a
power-law fashion. The upscattering of a power-law synchrotron spectrum by a power-law distribution 
of electrons leads to power-law inverse-Compton spectra (one of the input power-law suffices) of same
slopes as the synchrotron spectrum and to power-law light-curves of steeper decay (larger index $\alpha$),
as shown in table 2 of Panaitescu \& Vestrand (2012).

 The following are worth mentioning about the sample selection and power-law fits to their light-curves and spectra.

 Our selection of long-monitored LAT afterglows excludes afterglows whose flux exhibits a too fast
decay into the background, similar to a typical X-ray light-curve measured by Swift-XRT. With three exceptions 
(GRBs 090510, 110731, 130427), all other light-curves are well fit by a single power-law ($\chi^2_\nu \siml 1$) 
from their peaks to the last measurement above the background, 130427 being the only afterglow with a 
light-curve break, the other two exceptions displaying a single fluctuation above a power-law decaying flux,
after the burst for 090510, at the burst end for 110734. 

 Spectra are fit with a power-law starting from the lowest energy channel (100 MeV) and up to highest
energy for which the measurement is not more than $1\sigma$ above the power-law fit. 
Spectra are not always fit up to the highest energy channel (100 GeV), with the higher energy channels having 
sometimes only a flux $1\sigma$ upper limit, even if the resulting fit were acceptable. 
That is done for a few reasons, all related to the possible existence of a hard spectral component at higher 
energies (above few GeV).
One reason is that the power-law fit illustrates better the existence of that hard component. 
Second reason is that $\chi^2$ overestimates the fit quality when some channel fluxes lie predominantly 
on one side of the power-law fit. Third reason is to determine accurately the spectral slope $\beta$ at the lower 
energy channels, where most photons are, because that is the slope that should be related to the index $\alpha$
of the afterglow flux power-law decay. 

 Power-law best-fits are determined by minimizing 
\begin{equation}
  \chi^2 = \sum_i \left( \frac{F_{obs,i}-F_{pl,i}}{\sigma_i} \right)^2
\end{equation}
where $F_{obs,i}$ is the measured flux, $F_{pl,i}$ is the power-law model flux. 
For light-curves, $\sigma_i$ is the $1\sigma$ uncertainty of the photon flux $F_{obs,i} \equiv dN/dt$, 
but for spectra it is the Poisson statistics uncertainty of the model power-per-decade $F_{pl,i} \equiv 
E^2 (dN/dE)_{pl}$:
\begin{equation}
 \sigma \left(E^2 \frac{dN}{dE} \right) = E^2 \frac{ \sigma(C_{pl}) }{\Delta E A \Delta t} = 
                 E \sqrt{\frac{E^2(dN/dE)_{pl} }{\Delta E A \Delta t} }
\end{equation}
where $\sigma(C_{pl}) = \sqrt{C_{pl}}$ and $C_{pl} = (dN/dE) \Delta E A \Delta t$ is the model photon-count 
in channel of energy E, $\Delta E$ being the width of that channel, $\Delta t$ the spectrum integration-time, 
and $A$ the detector effective area ($A \Delta t$ is the full-exposure calculated by the {\sc gtexposure} tool).

 The reason for using the model flux uncertainty instead of the measurement uncertainty obtained with the
{\sc gtbin} tool is that the latter is calculated by summing over Poisson distributions of all integer 
average flux $C_o$, each distribution being weighted by the Poisson probability of measuring a flux $C_{obs}$ 
if the true flux were $C_o$.\footnote{
   One reason for why {\sc gtbin} calculates the measurement uncertainty in this way could be that, 
   for a non-detection $C_{obs}=0$, the Poisson distribution of average $C_0=0$ is a $\delta$ function, 
   whose dispersion $\sigma=0$ is un-usable. Another reason is that $C_{obs}$ photons being detected does not 
   imply that the measurement was drawn only from a Poisson distribution of average/peak value $C_{obs}$. } 
Consequently, for lower-energy channels (below 1 GeV), which often have few/several photons, 
the weighted-Poisson 
uncertainty calculated by {\sc gtbin} is $\sigma_{gtbin}(C_{obs}) \simeq \sqrt{C_{obs}}$ and Poisson statistics 
for the power-law model counts $C_{pl}$ (which is equal with the measured counts $C_{obs}$ when the best-fit is 
reached) is good enough for estimating the measurement uncertainty and for using in the $\chi^2$ calculation.
However, for higher-energy channels (above 1 GeV), which always have a few photons, $\sigma_{gtbin}(C_{obs}) > 
\sqrt{C_{obs}}$ and the use of measurement errors calculated by {\sc gtbin} would give little weight to the 
high-energy channels in determining the best-fit power-law. In that case, the use of model count uncertainty 
leads to a more accurate determination of the best-fit power-law index.

 $1\sigma$ uncertainties of the exponent of the best-fit power-law to light-curves and spectra are determined
from an increase $\Delta \chi^2 = 2.3$ above the minimal $\chi^2$ of the best-fit, i.e. the uncertainties are
calculated for the joint variation of both fit parameters (normalization and power-law index) These uncertainties 
are about 50 percent larger than those obtained by varying only one parameter of interest (the slope), which 
corresponds to $\Delta \chi^2 = 1$.

\section{Afterglow Light-Curves}

 The power-law fits to the light-curves and spectra of 24 well-monitored LAT afterglows are shown in Figures 2 and 3, 
and the best-fit power-law exponents are listed in Table 1. 
About two-thirds of light-curves show that the LAT light-curve peak occurred during the burst and the light-curves
do not show any deviation from a power-law at the end of the burst, indicating that the prompt emission mechanism 
does not contribute much to LAT emission above 100 MeV after the light-curve peak. This implies that, for a majority  
of afterglows in our sample, the LAT {\sl post-peak} emission is a spectral component brighter than the extrapolation 
to 100 MeV of the Band spectrum of the prompt/burst MeV emission. The joint spectral analysis of Fermi-GBM and 
LAT data during the burst phase shows that to be the case for GRB 080916C (Abdo et al 2009a), GRB 090510
(Ackermann et al 2010a, Giuliani et al 2010), GRB 090902B (Abdo et al 2009b), GRB 090926 (Ackermann et al 2011),
GRB 110731 (Ackermann et al 2013b), GRB 130427 (Ackermann et al 2014), GRB 131108 (Giuliani et al 2014),
but there is also a claim that the Band function describes well the entire GBM and LAT burst measurements over 
five decades in energy, for GRB 090217 (Ackermann et al 2010b).

\begin{figure*}
\centerline{\psfig{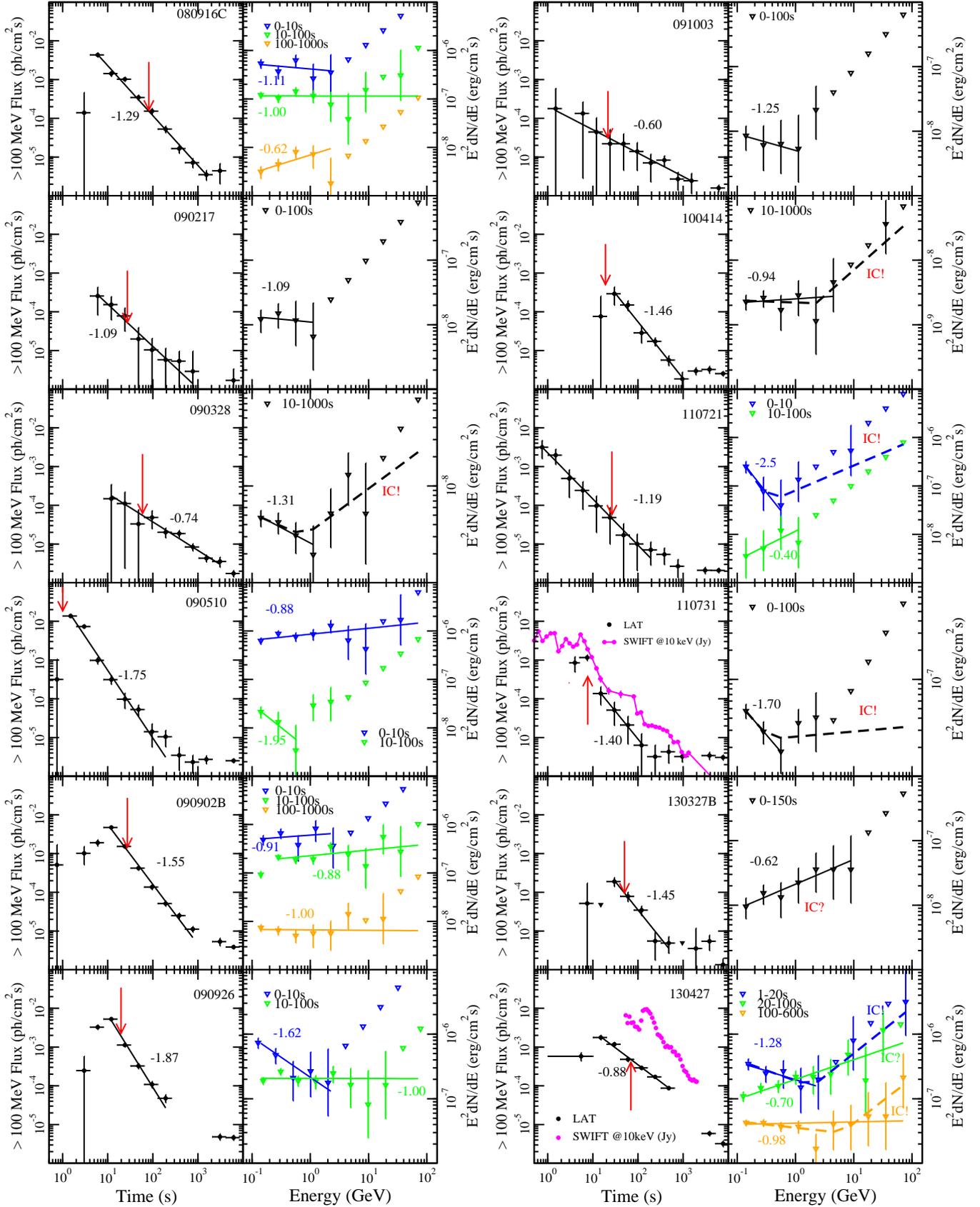}}
\figcaption{ Light-curves above 100 MeV (left panels) and 0.1--100 GeV power-per-decade spectra (right panels) 
     for 12 afterglows whose power-law flux decays have been monitored by Fermi-LAT over at least one decade in time. 
     {\sl Light-curves} time-bins extend over a factor two (0.3 dex).
     The red arrow indicates the end of the prompt-emission phase (the burst). 
      The 10 keV flux (in Jy) for Swift afterglows is also shown. 
      Numbers give the exponent $-\alpha$ of the power-law flux decay (Table 1).
     Triangles denote $1\sigma$ upper limits, and correspond to 1.9 photons. 
     (continued in next figure)
}
\end{figure*}

\begin{figure*}
\centerline{\psfig{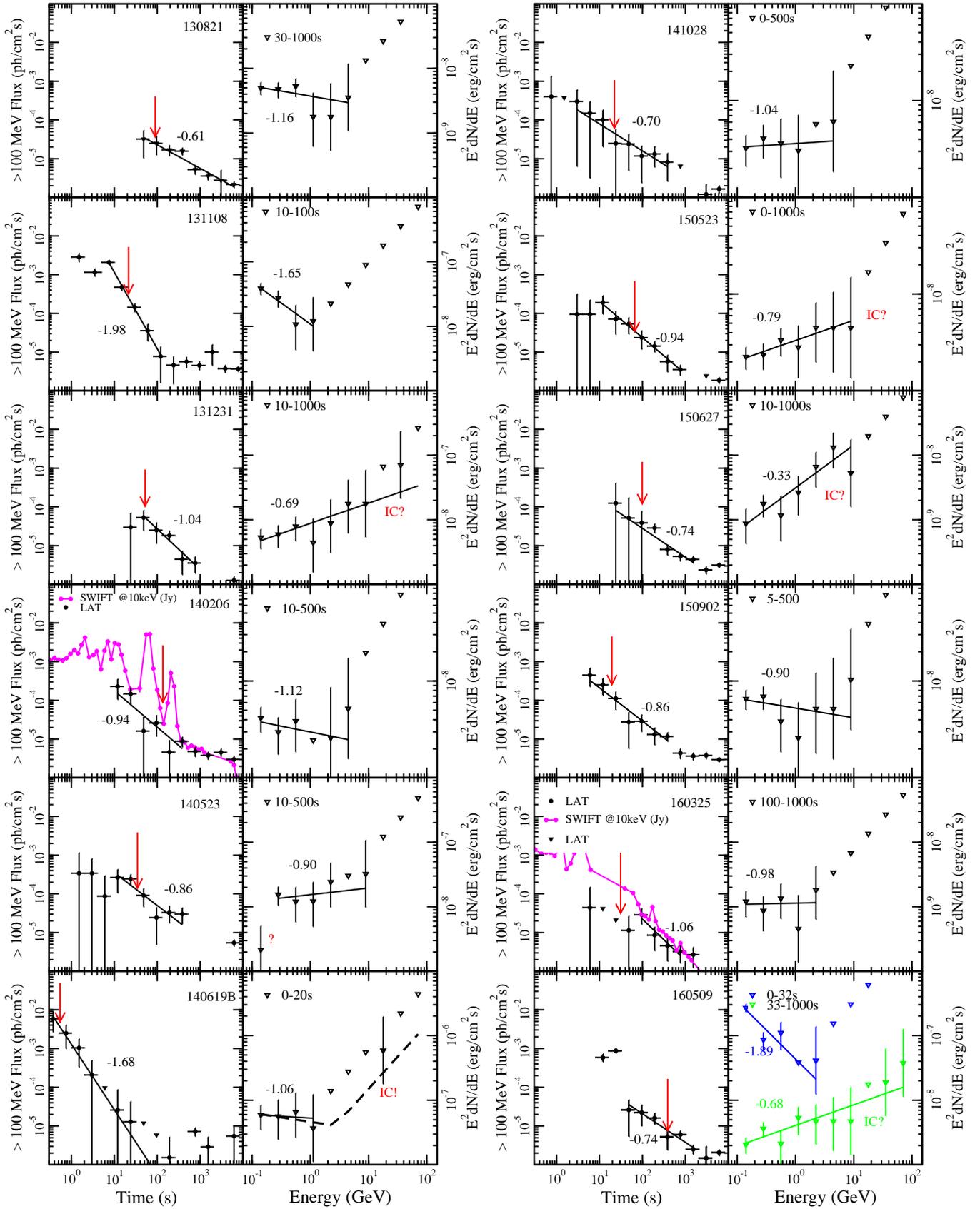}}
\figcaption{ Same as in Figure 2, for another 12 afterglows well-monitored by LAT.
     {\sl Spectral} energy-bins extend over a factor two.
     Single power-law fits are made starting from 100 MeV and up to where measurements can be fit within their error. 
     Numbers give the slope $-\beta$ of the power-law $F_E$ spectrum (Table 1).
     Triangles without error bars show $1\sigma$ upper limits. 
     (Because the {\sc gtbin} tool assigns a statistical uncertainty of 1.9 photons for any non-detection 
      and because the energy channel binning used here satisfies $\Delta E \propto E$, all non-detections follow 
      $E^2 (dN/dE) \propto E$).
     Dashed lines show broken power-law fits (labeled "IC!") for which the higher-energy harder component was found 
      to have a confidence level of at least 90 percent (Table 2). 
     Spectra harder than $F_E \propto E^{-1}$ (labeled "IC?") are also candidates for inverse-Compton emission.
     For 130427, we find good evidence for inverse-Compton above the spectral break at 1--20s and at 100--600s, 
      and also in the hard spectrum at 20--100s.
     Tam et al (2013) have identified the same hard spectral component at 0--20s, 138--750s, and after 3ks, and have
      proposed that it is inverse-Compton emission.  
 }
\end{figure*}

\begin{table*}[t]
 \caption{ Indices of power-law fits to the light-curves $dN/dt \propto t^{-\alpha}$ and spectra 
       $F_E = E(dN/dE) \propto E^{-\beta}$ of GRB afterglows that have been followed-up by Fermi-LAT over 
       at least a decade in time. 
       $1\sigma$ uncertainties of the best-fit slopes are for the joint variation of both fit-parameters 
       (slope and normalization).
       $\Delta t$ is the time interval over which the power-law fit was done.
       The power-law spectrum fit is over the lower energy channels where all measurements are within
       their $1\sigma$ uncertainty of the model flux (see Figures 2 and 3).
    }
\vspace*{4mm} 
\centerline{
\begin{tabular}{llllll|llllllllllllllllll}
   \hline
        & \multicolumn{2}{c}{\underline{\sc Light-curve}} & \multicolumn{2}{c}{\underline{\sc Spectrum}} & & &
        & \multicolumn{2}{c}{\underline{\sc Light-curve}} & \multicolumn{2}{c}{\underline{\sc Spectrum}} \\
   GRB  &  $\Delta t$(s)  &  $\alpha(\sigma_\alpha)$  &  $\Delta t$(s)  &  $\beta(\sigma_\beta)$    &  & &
   GRB  &  $\Delta t$(s)  &  $\alpha(\sigma_\alpha)$  &  $\Delta t$(s)  &  $\beta(\sigma_\beta)$         \\
  \hline
   080916C &  4-2000 & 1.29(.05) &  1-1000 & 1.01(.09) &&& 090217  &  4-1000 & 1.09(.42) &  1-100  & 1.09(.47)  \\
   090328  & 10-4000 & 0.74(.13) & 10-1000 & 1.31(.28) &&& 090510  &  1-130  & 1.77(.11) &  1-100  & 0.91(.09)  \\
   090902B &  10-700 & 1.60(.08) &  1-700  & 0.95(.08) &&& 090926  &  10-250 & 1.87(.13) &  1-100  & 1.11(.09)  \\
   091003  &  1-2000 & 0.60(.21) &  1-100  & 1.25(.53) &&& 100414  & 25-1000 & 1.46(.21) & 10-1000 & 1.06(.21)  \\
   110721  & 0.5-200 & 1.19(.22) &  1-100  & 1.35(.33) &&& 110731  &  10-150 & 1.43(.65) &  1-100  & 1.70(.35)  \\
   130327B &  20-500 & 1.45(.26) & 10-160  & 0.62(.13) &&& 130427  &  10-600 & 0.88(.06) & 20-600  & 1.17(.06)  \\
   130821  & 30-7000 & 0.61(.12) &  1-600  & 1.16(.23) &&& 131108  &  5-150  & 1.98(.22) & 10-100  & 1.65(.35)  \\
   131231  & 30-1000 & 1.04(.34) &  10-300 & 0.69(.16) &&& 140206  &  10-400 & 0.94(.38) & 10-400  & 1.12(.26)  \\
   140523  & 10-500  & 0.85(.36) &  10-500 & 0.90(.28) &&& 140619B &  0.3-30 & 1.66(.46) &  1-20   & 1.06(.50)  \\
   141028  & 2-400   & 0.70(.50) &  1-400  & 0.94(.29) &&& 150523  & 10-1000 & 0.94(.18) & 1-1000  & 0.79(.17)  \\
   150627  & 20-2000 & 0.74(.32) &  1-1000 & 0.34(.17) &&& 150902  &  5-400  & 0.86(.24) & 1-500   & 1.10(.19)  \\
   160325  & 90-1000 & 1.06(.70) &  90-1000& 0.98(.30) &&& 160509  & 30-1000 & 0.74(.20) &  30-1000& 0.68(.12)  \\        
  \hline 
\end{tabular}
}
\end{table*}

\begin{table*}
 \caption{
    Results of the $F$-test for the statistical significance of the high-energy harder spectral component. 
    $\Delta t$ is the spectrum integration time, $\chi^2_{spl}$ and $\chi^2_{bpl}$ are for the single power-law and 
    broken power-law best-fits to ten-energies channels. $E_*$ is the best-fit value for the dip energy, above
    which the spectrum hardens, and its $1\sigma$ range was determined by a $\Delta \chi^2 =1$ increase around the best fit.
    }
\vspace*{4mm}
\centerline{\small
\begin{tabular}{lllllll|lllllllllllllllll}
  \hline
   GRB &$\Delta t$&$\chi^2_{spl}$&$\chi^2_{bpl}$& BPL &$E_*$&$\sigma(E_*)$& GRB   &$\Delta t$&$\chi^2_{spl}$&$\chi^2_{bpl}$&BPL  &$E_*$&$\sigma(E_*)$ \\  
       &   (s)    &   (8 df)    &  (6 df)     & prob&(GeV)& (GeV)   &       &  (s)     &  (8 df)     &  (6 df)     &prob & (GeV)& (GeV)   \\ 
  \hline
 090328& 10-1000  &   10.4       &   6.5      & 0.90&0.8  & 0.4-1.4 & 100414& 10-1000  &   12.1      &   6.0       & 0.95& 2.3  & 1.2-14  \\ 
 110721& 0-10     &   13.6       &   3.6      &0.994&0.4  & 0.2-0.6 & 110731&  0-100   &   7.8       &  4.9        &0.90 & 0.4  & 0.2-0.7  \\
 130427&  1-20    &   14.8       &   4.3      &0.992&1.9  & 0.9-5.1 &140619B&  0-20    &   6.1       &  3.4        &0.93 & 2.7  & 0.8-8.5  \\
  \hline
\end{tabular}
}
\end{table*}

 All fits are plotted together in Figure 4 (left panel), which suggests a correlation between the afterglow 
brightness and its decay: brighter afterglows decay faster than dimmer afterglows.
For the fast-decaying afterglows with $\alpha > 1$, most of their energy release occurs at the
light-curve peak, while for slow-decaying afterglows with $\alpha < 1$, most of their energy release
occurs well after the peak. Part of the above correlation is a selection effect because retaining only 
 afterglows that have been long monitored excludes dimmer afterglows with a fast fall-off, as they get below 
the background in less than a decade in time since first detection. 
However, the rest of that correlation is real: bright afterglows with a slow dimming rate are absent, with the 
exception of 130427. Given that the brightness distribution at 10 s has a width of 2 dex, which is at least 
1 dex more than what the spread by a factor 2-3 in redshift can induce, the brightness-decay correlation implies 
a luminosity-decay correlation. 

  A similar correlation was observed for the optical emission of afterglows displaying an early fast-rise to a 
peak at about 100s (Panaitescu \& Vestrand 2008). The energetic output of the brighter/fast-falling optical 
light-curves was found to be better correlated with the GRB output than for the dimmer/slowly-decaying optical 
light-curves (Panaitescu \& Vestrand 2011), which suggests that the above dichotomy (bright+fast decays and
dim+ slow-decays) of optical afterglows is due to how the outflow energy is deposited in the external shock:
impulsive outflows (released on a short timescale) lead to bright optical afterglows 
(because all outflow's energy was quickly deposited in the shock) while extended outflows yield dimmer optical 
afterglows (because only a fraction of the entire outflow energy was in the afterglow shock at early times)
and to a slower-decaying optical light-curves (because of the long-lived energy injection in the external shock).  

 We extend the above conclusion reached with a sample of 33 optical afterglows to the current 24 LAT afterglows,
and attribute the bright/faster-decaying afterglows to an impulsive release of the relativistic ejecta and the 
dimmer/slower-decaying afterglows to an extended ejecta release. 

 For more than half of the light-curves of Figures 2 and 3, the 1--50 s LAT light-curve peak occurs during the prompt 
GRB emission. Optical light-curves also display peaks, but at later times (50--1000s). The difference in peaks
epochs is due to $1)$ a selection bias against late LAT peaks, which may be too dim to allow a long-monitoring of the 
afterglow, and $2)$ an observational bias against early optical peaks, which may be missed due to the latency 
of robotic optical telescopes to respond to GRB trigger notices.
A minority of Swift X-ray afterglow light-curves also display a peak after 100 s (Panaitescu, Vestrand, Wozniak 2013), 
because peaks occurring earlier are overshined by the prompt emission.

 Therefore, LAT afterglow observations are best to catch the earlier light-curve peaks. 
As for the optical and X-ray light-curve peaks, we propose that the most likely reason for the LAT peaks is the onset 
of the forward-shock deceleration, although other explanations are possible: $1)$ light-curve peaks at any observing 
frequency arise if the direction toward the observer is not initially within the jet opening, when the jet has
decelerated enough that the direction toward the observer has entered the relativistically-beamed emission cone, 
$2)$ LAT light-curve peaks could be due to the emission above 100 MeV becoming optically-thin to pair-formation.

 If light-curve peaks arise from the onset of deceleration, then the peak epoch is
\begin{equation}
 t_p = (z+1) \frac{R_d}{2c \Gamma_0^2} = 16\frac{z+1}{3} \h1 \left(\h1  \frac{E_k(t_p)}{10^{53}\, erg} 
       \frac{1\,cm^{-3}}{n} \right)^{\h1 1/3} \h1 \h1 \left( \frac{\Gamma_0}{400} \right)^{\h1 -8/3} \h1 \h1 s
\label{tp}
\end{equation}
where $E_k$ is the isotropic-equivalent kinetic energy of forward-shock at time $t_p$, $\Gamma_0$ is the ejecta 
initial Lorentz factor, $z$ is the afterglow redshift, and $n$ is the external medium proton density at the 
deceleration radius $R_d$, where the forward-shock has swept-up a mass $\Gamma_0$ smaller than that of the ejecta. 
Panaitescu et al (2013) have shown that variations in the ambient medium density among afterglows lead to a 
peak-flux $F_p$ -- peak-epoch $t_p$ anticorrelation consistent with that measured for 30 optical peaks: $F_p \propto 
t_p^{-(2-3)}$. Variations in $\Gamma_0$ lead to an anticorrelation of a smaller slope, while those in $E_k$ produce
a correlation. Performing the same test with LAT light-curves requires more than the seven light-curve peaks (shown 
in Figures 2 and 3) of afterglows with known redshift.

 Continuing with the generally-accepted identification of the LAT afterglow emission with the forward-shock's
(e.g. Kumar \& Zhang 2015), we compare in Figure 4 (right panel) the temporal and spectral power-law indices 
$(\alpha,\beta)$ with the expectations for that model. The forward-shock model predicts a closure-relation
$\alpha = k\beta + c$ with $k$ depending on the shock geometry (spherical, conical jet, sideways spreading jet)
and radiation mechanism (synchrotron, inverse-Compton) dominant at the observing frequency and $c$ depending on 
the radiation mechanism, radial distribution/stratification of the ambient medium (homogeneous, wind-like),
and location of the observing frequency relative to the breaks (peak, cooling) of the afterglow spectrum 
(see table 2 of Panaitescu \& Vestrand 2012 for all possible values of $c$ and $k$, including also the effect 
of energy injection in the blast-wave). Thus, the $\alpha-\beta$ correlation expected for the forward-shock
is that softer afterglows should decay faster.
 
 The linear correlation coefficient of the decay indices and spectral slopes shown in Figure 4 is $r = 0.29 \pm 0.26$.
A similarly weak correlation is manifested by Swift X-ray afterglows at early times (figure 4 of Panaitescu 2007).
The $r=0.29$ corresponds to a 20 percent probability of this correlation occurring by chance, and the weakness
of this correlation could be due to $1)$ more than one forward-shock sub-model occurring in LAT afterglows and 
to $2)$ the limited baseline of both indices. The large $\sigma_r \siml r$ reflects the substantial uncertainties 
of one or both indices and makes the $\alpha-\beta$ correlation only marginally significant. 

 Given the large uncertainty in the temporal and spectral indices shown in Figure 4, nearly all LAT afterglows 
are consistent with a forward-shock without energy injection. Still, half of the slowly-decaying LAT afterglows 
($\alpha < 1$) display decays slower than for the slowest decay expected from an adiabatic forward-shock 
(synchrotron emission with synchrotron cooling energy below 100 MeV). Thus, Figure 4 offers some evidence 
for a sustained energy-injection in the forward-shock.
 
 Last note about the light-curves shown in Figures 2 and 3 is that the LAT flux decay appears to be consistent
with the contemporaneous light-curve measured by Swift-XRT at 1--10 keV in four cases, to the extent that the
X-ray flares present in all those four afterglows can be ignored. We note that six LAT afterglows display a 
flux-decay slower than $dN/dt \propto t^{-3/4}$, which would be the LAT counterpart of the X-ray plateaus 
seen by Swift-XRT in a many afterglows (e.g. Nousek et al 2006). 

\begin{figure*}
\centerline{\psfig{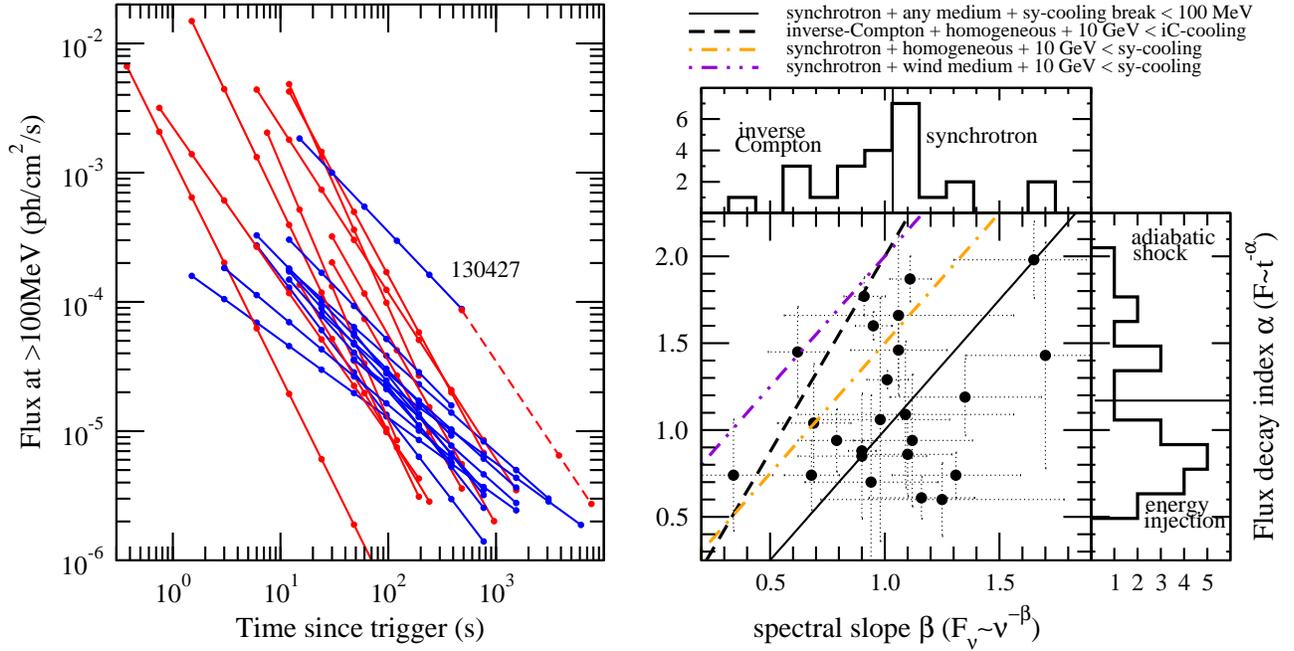}}
\figcaption{ {\sl Left} panel: 24 LAT afterglow light-curves, with fast decayers ($\alpha < -1.2$) shown in red
      and slow decayers ($\alpha > -1.2$) in blue. 130427 is the only case with a clear light-curve break/steepening,
      and its break at 4 ks is chromatic, not seen at any other wavelength (optical or X-ray).
      {\sl Right} panel: flux decay-index $\alpha$ vs the index $\beta$ of the power-law energy spectrum. 
      Dots show 1$\sigma$ uncertainties. 
      Lines indicate the $\alpha-\beta$ relation expected for the synchrotron and inverse-Compton emissions
      from an adiabatic forward-shock interacting with a homogeneous medium or a wind-like one, with the
      LAT range being located below or above the spectral cooling-break corresponding to the electrons whose 
      radiative cooling-time equals the timescale over which a new generation is produced (i.e. the
      afterglow dynamical timescale).
      Lines in the histograms of indices $\alpha$ and $\beta$ show the corresponding average values.
      Faster-decaying light-curves may be associated with an impulsive ejecta release, while slower-decaying
      afterglows with energy injection in the forward-shock.
      Harder spectra could be inverse-Compton dominating the LAT emission, while softer spectra are more likely
      to be synchrotron emission.
  }
\end{figure*}

\vspace*{2mm}
\section{Afterglow Spectra}

\subsection{High-energy hard spectral component from inverse-Compton}

 Out of the 24 LAT afterglows considered here, we find evidence (above 90 percent confidence level) of a harder 
component at higher-energies (above a dip or ankle energy) in six cases, which are listed in Table 2. 
With $\Delta \chi^2 = \chi^2_{spl} - \chi^2_{bpl}$ the reduction in $\chi^2$ produced by the addition of 
another power-law at higher energies and the reduced chi-square $\chi^2_\nu = \chi^2_{bpl}/\nu$  of the broken 
power-law fit, the probability that the $\Delta \chi^2$ reduction is accidental follows the $F$-distribution 
$P(F;1,\nu)$, where $F = \Delta \chi^2/\chi^2_\nu$ and $\nu = 6$ is the number of degrees of freedom for the 
broken power-law fit (10-channel fluxes minus four model parameters). Then, $1-P(F;1,6)$ is the probability 
for the hard spectral component being real (listed in Table 2).

 If both spectral components are from the same shock, then it seems natural to conclude that the lower-energy 
component is synchrotron and the higher-energy component is inverse-Compton. At such high energies, the afterglow
synchrotron emission is most likely measured above the cooling break, where the spectral slope of $F_E$ is 
$\beta_l = p/2$, with $p$ being the index of the power-law distribution of electrons with energy in the shock. 
For the higher-energy component to be harder than the synchrotron emission at lower energies, the inverse-Compton 
spectrum must be measured by LAT below the upscattered cooling break, where the spectral slope is $\beta_h = (p-1)/2$. 
Therefore, in this interpretation of the spectral dip of the six afterglows of Table 2, one expects a spectral
hardening across that dip by $\beta_h - \beta_l = 1/2$, which is compatible with (or less than) the hardening 
$\Delta \beta$ displayed by the spectra of Figures 2 and 3, but not clearly inconsistent with them because the 
high-energy spectral slope is very uncertain.

 If this synchrotron self-Compton interpretation of the LAT spectra with a break is correct, and if the LAT emission
arises in the forward-shock, then the dip energy $E_*$ can be calculated by equating the synchrotron and inverse-Compton 
fluxes: $F_{sy}(E_*) = F_{iC}(E_*)$, leading to $E_* = E_c^{(sy)}/Y^2$ for $\beta=1$, with $E_c^{(sy)}$ the cooling-break 
energy of the synchrotron spectrum and $Y < 1$ the Compton parameter. From here, one obtains
\begin{equation}
 E_* \h1 =  1  \left( \frac{E_k}{10^{53}\, erg}\right)^{\h1 -1} \h1 \h1 \left( \frac{\varepsilon_e}{10^{-3}} \right)^{\h1 -2} 
   \h1  \left( \frac{\varepsilon_B}{10^{-5}} \frac{n}{1\, cm^{-3}} \right)^{\h1 -1.5} \h1 \h1 {\rm GeV}
\end{equation}
for a homogeneous external medium of proton density $n$ (in this case, $E_*$ is time-independent) and
\begin{equation}
 E_* = 5 \left( \frac{E_k}{10^{53}\, erg}\right)^{\h1 0.5} \h1 \h1 \left( \frac{\varepsilon_e}{10^{-3}} \right)^{\h1 -2} 
     \h1  \left( \frac{10^{-9}}{\varepsilon_B} \frac{t}{100\,s} \right)^{\h1 1.5}  {\rm GeV}
\end{equation}
for the $n \propto r^{-2}$ wind of a Wolf-Rayet star (as the GRB progenitor), with $E_k$ the forward-shock kinetic
energy, $\varepsilon_e$ and $\varepsilon_B$ the post-shock fractional energies in electrons and magnetic field, 
respectively, and redshift $z=1$ was assumed.\footnote{
 The wind case requires a very weak magnetic field in the forward-shock, especially if $\varepsilon_e \simg 0.01$,
 as obtained from modeling the afterglow broadband emission.}
The above equations indicate that, if the shock is adiabatic (no energy-injection) and if the two micro-parameters
do not evolve, then the dip energy $E_*$ is constant for a homogeneous medium and increases for a wind-like medium.
The multi-epoch spectra of Figures 2 and 3 provide a very weak support for a constant or a decreasing dip energy $E_*$. 

 Also from the above two equations, it is evident that $E_*$ may fall below 100 MeV, in which case all LAT emission 
would be inverse-Compton. Because the synchrotron spectrum above 100 MeV can only be softer than the inverse-Compton,
it seems natural to correlate the hardness of LAT spectra with the afterglow emission process and attribute
spectra softer (harder) than $\beta = 1$ to synchrotron (inverse-Compton). Afterglows 130327B, 131231, 150523, 150627,
and 160509, as well as 130427 at the second epoch shown in Figure 2, display a hard power-law over the entire LAT range; 
we suggest that the LAT emission of those afterglows was inverse-Compton.

\clearpage
\subsection{Spectral steepening due to photon-photon absorption}

 If the high-energy afterglow emission is optically-thick to pair-formation above an energy $E_\tau$ that falls
within the LAT range, then a steepening of the spectrum above that "cut-off" energy should be observed. 
This attenuation is most likely to be seen early during the afterglow,
when the source is smaller and the pair-formation opacity larger. Assuming that the source Lorentz factor is
of at least a few hundreds, an observed 1 GeV photon will form pairs with photons of threshold-energy above 10 MeV. 
While there could be a measurable output above 10 MeV from the sub-MeV burst mechanism, those target photons
are dimmer than the emission above 10 MeV of the brighter afterglows of interest here (see below),
thus we will restrict to pair-formation of afterglow photons on other afterglow photons.

 For the pair-formation in an un-decelerated source, which is applicable for the external-shock at the peak of 
its light-curve, we (i.e. I, Panaitescu 2015) have shown that the optical-thickness is
\begin{equation}
  \tau (E) \h1 = \h1 0.46 \left( \frac{z+1}{3}\right)^{\h1 6}  \h1 \frac{\Phi}{10^{-2}\cm2} 
                \h1 \left( \frac{\Gamma}{200} \right)^{\h1 \h1 -6} 
               \h1 \h1 \left( \frac{t}{10\,s} \right)^{\h1 \h1 -2} 
               \h1 \h1 \left( \frac{E}{1\, GeV}  \right)^{\h1 \beta}
\end{equation}
where $\Phi$ is the photon fluence, $\Gamma$ is the source Lorentz factor, both functions of the observer time $t$,
and $\beta$ is the afterglow energy-spectrum slope. 
The cut-off energy $E_\tau$ defined by $\tau(E_\tau)=1$ is therefore
\begin{equation}
  \h1 \h1 \h1  E_\tau = 1.0 \left( \frac{z+1}{3}\right)^{\h1 -6} \h1 \h1 \left( \frac{\Phi}{10^{-2}\cm2} \right)^{\h1 -1}
        \h1 \h1 \left( \frac{\Gamma}{175} \right)^{\h1 6} \h1 \left( \frac{t}{10\,s} \right)^{\h1 2} {\rm GeV}
\label{Ec1}
\end{equation}
assuming $\beta = 1$. Given that, for the expected blast-wave energetics (comparable to the burst output) and 
ambient medium density, afterglows could be even more relativistic than $\Gamma = 200$ at $t = 10s$, the above 
equation shows that a cut-off energy in the LAT range could be seen only for the brightest afterglows in Table 1.
If $E_\tau$ is measured at some time $t$, for an afterglow of fluence $\Phi(t) \sim t F(t)$, then the source 
Lorentz factor can be determined:
\begin{equation}
  \Gamma_\tau = 175\; \frac{z+1}{3} \left( \frac{\Phi}{10^{-2}\cm2} \frac{E_\tau}{1\, GeV} \right)^{1/6}
        \left( \frac{t}{10\,s} \right)^{1/3} 
\label{Gtau}
\end{equation}

\begin{figure*}
\centerline{\psfig{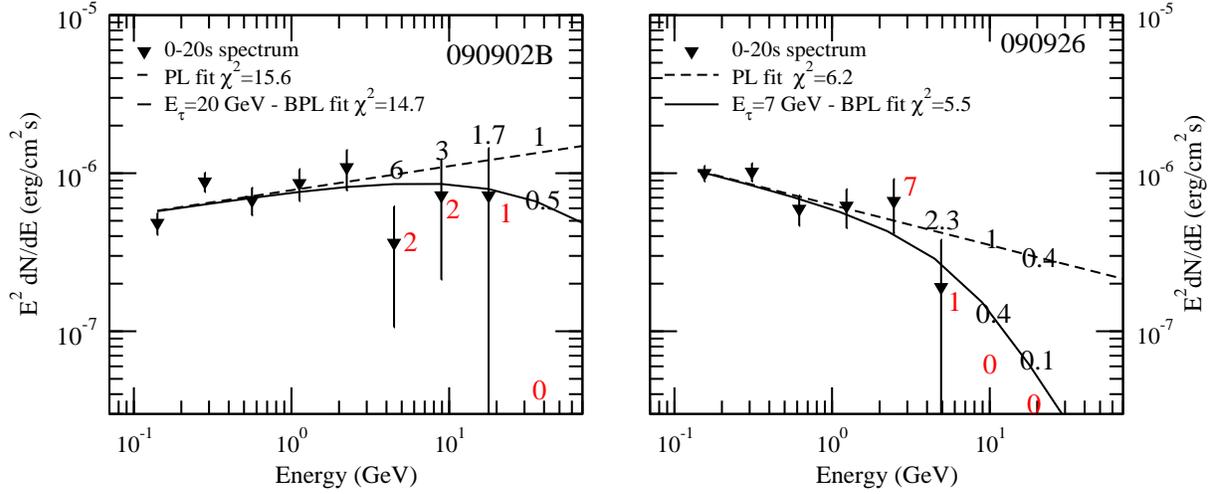}}
\figcaption{ Early spectra of the two brightest (in fluence, around light-curve peak) afterglows, showing
  power-law fits (dashed line) and best-fits including attenuation for pair-formation, with the "cut-off" 
  energies in legends. The cut-off is not exponential, but a transition to a steeper power-law (see text). 
  Black numbers next to lines indicate the model number of photons in that channel; red show the number of LAT photons.
  The reduction in $\chi^2$ obtained by the addition of the cut-off corresponds to a chance probability of
  obtaining that fit improvement of 0.54 for 090902B and 0.35 for 090926.
  Both break energies are poorly constrained: a $\Delta \chi^2 =1$ increase of the best-fit $\chi^2$ (in legend)
  sets only lower limits on the break energy: $E_\tau > 8$ GeV for 090902B and and $E_\tau > 4$ GeV for 090926.
  Spectra integrated over shorter durations have fewer photons and higher Poisson
  fluctuations, and do not provide better evidence for a spectral cut-off.  }
\end{figure*}

 For the attenuation of a photon-front with itself, the fraction of escaping photons is $(1-e^{-\tau})/\tau$,
where $\tau$ is the above optical-thickness of the entire front (for a photon passing through it), thus an
intrinsic $F_0 (E) \propto E^{-\beta}$ emission spectrum becomes an attenuated one:
\begin{equation}
 F(E) \propto \frac{1-\exp[-\tau(E)]}{\tau(E)} E^{-\beta} \;, \tau(E) = \left( \frac{E}{E_\tau} \right)^\beta
\label{Fatt}
\end{equation} 
therefore
\begin{equation}
 F(E) \propto \left\{ \begin{array}{ll} E^{-\beta} & E \ll E_\tau \; \left[\tau(E) \ll 1 \right]  \\ 
              \frac{\displaystyle E^{-\beta}}{\displaystyle \tau(E)} \propto E^{-2\beta} & 
           E_\tau \ll E \; \left[\tau(E) \gg 1 \right] \end{array}  \right.
\label{bpl1}
\end{equation}
Thus, the cut-off due to photon-photon absorption is not the exponential reduction expected for a photon passing 
through the entire front, but a break to a steeper power-law (hence $E_\tau$ is just a "break" energy), owing to 
those photons produced in the outer one optical-depth, which escape unabsorbed.
 
 Equation (\ref{Ec1}) shows that the photon-attenuation break-energy $E_\tau$ increases with the light-curve peak 
epoch and with decreasing peak fluence. Given that there are fewer photons at higher energies, where a spectral 
break can be lost in statistical fluctuations, the best chance to measure $E_\tau$ is for fluence-bright LAT light-curve 
peaks that occur the earliest. 

 In our sample, we find only two early afterglow spectra that display a softening break,
as shown in Figure 5. Their spectra were fit with both a power-law and the attenuated power-law given in
equation (\ref{Fatt}). An $F$-test indicates that the break in the 0-20s spectrum of 090902B is not statistically 
significant, while that of 090926 is just marginal, at $1\sigma$ confidence level.
 That is not surprising, considering that the fluxes measured by LAT in the higher energy channels fall short 
by only two photons relative to the unabsorbed spectrum (the extension to high energy of the power-law fit to 
the lower-energy fluxes). 
 For the best-fit break energies, $E_\tau=20$ GeV for 090902B and $E_\tau = 7$ GeV for 090926, and for their 
peak fluences and redshifts, equation (\ref{Gtau}) yields about the same source Lorentz factor $\Gamma_\tau \simeq 
400$ at 10 s. Taking into account the large uncertainties in determining those break energies, only lower limits
can be set: $\Gamma > 330$ for 090902B and $\Gamma > 350$ for 090926.  \footnote{  
   Ackermann et al (2011) also concluded that the LAT data offer only a weak evidence for a break in the
 3-22s spectrum of 090926, and found $E_\tau = 0.2$ GeV for a broken power-law fit and $E_\tau = 1.4$ GeV for an
 exponential cut-off; using the burst variability timescale $\Delta t=0.15$ s instead of the peak epoch
 $t \simeq 13$ s, they inferred a higher Lorentz factor $\Gamma_\tau = 720$.}

 The attenuated spectra of Figure 5 indicate that a strong evidence for a power-law softening
at higher energies requires that each power-law branch is measured over a decade in photon energy and that 
the highest energy channel has at least one photon. 
With a $\Delta E \simeq E$ energy-binning of the spectrum, the number of photons per energy-channel follows 
$N(E) \propto E^{-\beta}$, for an unabsorbed energy-spectrum $F_0(E) \propto E^{-\beta}$. Starting with the 
minimal $N(10\, GeV) = 1$ photon in the 10 GeV channel, and going toward lower energies with an attenuated 
$F(E) \propto E^{-2\beta}$ spectrum, requires $N(E_\tau) = 10^{2\beta}$ photons at $E_\tau = 1$ GeV. From here, 
the unabsorbed spectrum would have $N(0.1\, GeV) = 10^\beta N(1\, GeV) = 10^{3\beta}$ photons which, for 
an average LAT effective-area of 9000 cm$^2$ for on-axis photons, implies a fluence $\Phi = 10^{3\beta-4}\,\cm2$.
Thus, for the average spectral slope $\beta = 1$, an afterglow spectrum will show good evidence 
for the photon-attenuation broken power-law with a break at 1 GeV if its fluence at the light-curve peak is 
$\Phi = 0.1\,\cm2$. Ten afterglows shown in Figures 2 and 3 have spectra harder than $\beta = 1$ and
lower peak fluences, with measurements spanning 2-3 decades in energy, but do not show a spectral softening; 
instead, some display a harder component at higher energies. For those few afterglows with known redshift, 
equation (\ref{Gtau}) with $E_\tau > 10-100$ GeV can be used to set a lower limit on their Lorentz factor 
at the light-curve peak.

\subsection{Steepening of the synchrotron spectrum}

 If the electron acceleration mechanism is first order Fermi and if the magnetic field is perpendicular
to the shock, then the electron gyration time $T_L = 2\pi/\omega_L = 2\pi \gamma m_e c^2/eB$ in the 
down-stream fluid underestimates the time that the electron spends in a high magnetic field $B$, while 
bouncing off an increasing $B$ or on regions of uniform $B$ before returning upstream and, then, being 
deflected in the upstream region and caught by the shock for another energy-gaining shock-crossing. 
However, if the magnetic field is parallel to the shock, then the Larmor period $T_L$ is a better approximation 
for the duration during which the electron losses its energy, because the electron exits the shock after a
half-orbit. Thus, the gyration timescale $T_L$, which increases with the electron energy $\gamma m_e c^2$, 
brackets the time that an electron spends in the shocked fluid.
 
 Electrons that lose their energy on a synchrotron-cooling timescale (assuming that inverse-Compton losses are weaker, 
$Y < 1$) $T_{sy} = \gamma m_e c^2/ P_{sy}$, with the synchrotron power $P_{sy} = (1/6\pi) \sigma_e c (B\gamma)^2$ 
for a pitch-angle $\theta = \pi/2$, shorter than the time $\sim T_L (\gamma)$ that they spend in the shocked fluid 
will not exit the shock and be re-accelerated, thus the condition $T_{sy} (\gamma) = T_L (\gamma)$ sets an
approximate upper-limit on the energy that electrons can acquire through first-order Fermi acceleration:  
\begin{equation}
  \gamma_s(\p2) = \left( \frac{3e}{\sigma_e B} \right)^{1/2} = \frac{24\, TeV/m_e c^2}{\sqrt{B}}
\label{gmax}
\end{equation}
where $\sigma_e = (8\pi/3) r_e^2$ is the electron scattering cross-section, with $r_e = e^2/m_e c^2$ being the 
classical electron radius. Electrons of energy $\gamma_s(\p2)$ radiate synchrotron photons at a characteristic 
frequency 
\begin{equation}
  \omega_c [\gamma_s(\p2)]= \frac{3e}{2 m_e c} [\gamma_s(\p2)]^2 B = \frac{27}{16 \pi} \frac{c}{r_e}
\end{equation}
that is independent of the magnetic field strength $B$. The corresponding synchrotron-spectrum averaged energy,
after a relativistic boost $\Gamma$ and at a redshift $z$, is
\begin{equation}
 E_s = 1.7\, \frac{3}{z+1} \frac{\Gamma}{100} \, {\rm GeV}
\label{Ec2}
\end{equation}

 If inverse-Compton is the emission process dominant at LAT energies, then the maximal upscattered photon energy 
corresponding to the maximal electron energy given in equation (\ref{gmax}) is well above the LAT range.
However, if the LAT emission process is synchrotron, then the spectral cut-off associated with the maximal 
electron energy could fall within the LAT energy range (equation \ref{Ec2}), and its measurement determines 
the source Lorentz factor
\begin{equation}
 \Gamma_s \simg 60 \frac{z+1}{3} \frac{E_s}{1\, GeV}
\label{Gs}
\end{equation}
Being softer, the emission of 090926 (Figure 5) is more likely to be synchrotron. If its spectral break arises from
the upper-limit on the energy that electrons acquire through shock-acceleration ($E_s > 4$ GeV), then $\Gamma > 250$ 
for 090926 at about 10 s. If also synchrotron emission, the break of 090902B ($E_s > 8$ GeV) implies $\Gamma > 440$.

 Given that the synchrotron power depends on the electron pitch-angle, $P_{sy} \propto \sin^2 \theta$, the equality 
of the gyration period with the synchrotron cooling timescale leads to $\gamma_s (\theta) = \gamma_s (\p2)/\sin \theta$. 
Thus, electrons that enter the shock with an energy $\gamma > \gamma_s(\p2)$ will lose all their energy in the 
downstream region if their pitch-angle satisfies $\theta > \theta_c (\gamma) \equiv \sin^{-1}[\gamma_s(\p2)/\gamma]$ 
and will not cross the shock again and will not be re-accelerated. Conversely, electrons with pitch-angles satisfying 
$\theta < \theta_c(\gamma)$ will not lose all their energy before going back to the upstream region, thus such 
electrons could be accelerated to energies higher than their current $\gamma$. If the pitch-angle distribution is
isotropic, then a fraction $f(\gamma) = 1- \cos \theta_c(\gamma) = 1 - \sqrt{1 - [\gamma_s(\p2)/\gamma]^2}$ of
the current generation of electrons with $\gamma > \gamma_s (\p2)$ will go to the next shock-crossing.
Together with that the escape of electrons in the downstream region and in the loss-cone in the upstream lead 
to a power-law distribution with energy ($dN_e/d\gamma \propto \gamma^{-p}$), the radiative cooling for electrons
with $\gamma > \gamma_s(\p2)$ means that, at $\gamma_s (\p2)$, the electron distribution steepens
from $dN_e/d\gamma \propto \gamma^{-p}$ to $dN_e/d\gamma \propto \gamma^{-p} \{1-\sqrt{1-[\gamma_s(\p2)/\gamma]^2}\}$,
which is $dN_e/d\gamma \propto \gamma^{-(p+2)}$ for $\gamma \gg \gamma_s(\p2)$. Because the slope of the synchrotron
spectrum is $\beta = (p-1)/2$, a steepening of the electron distribution by $\Delta p = 2$ corresponds to a spectral
softening by $\Delta \beta = 1$. Thus, the synchrotron spectrum including the effect of electron radiative cooling 
during its acceleration (i.e. during one gyration timescale) is 
\begin{equation}
 F(E) \propto \left\{ \begin{array}{ll} E^{-\beta} & E \ll E_s \\ E^{-(\beta+1)} & E_s \ll E \end{array}  \right.
\label{bpl2}
\end{equation}
with $E_s$ given in equation (\ref{Ec2}).
For the average spectral slope $\beta = 1$, the steepening of the synchrotron spectrum at $E_s$ due to electron 
cooling during one gyration period \footnote{
 Not to be confused with the softening of the synchrotron spectrum by $\Delta \beta = 1/2$ due to electron cooling 
 during one dynamical timescale, which is at a much lower energy than $E_s$. }
given in equation (\ref{bpl2}) is the same as the softening at $E_\tau$ due to photon-photon absorption given
in equation (\ref{bpl1}). 

 Thus, LAT measurements of afterglow spectral breaks are unlikely to distinguish between the above two 
processes/mechanisms.

\vspace*{4mm}
\section{Conclusions}

 We have presented the light-curves and spectra of 24 Fermi-LAT afterglows observed in the first 7$\twothirds$ years of operation
(Figures 2 and 3).  The only selection criterion was that the light-curve was monitored over at least a decade in time. 
The reason for that is to allow the determination of the flux power-law decay and compare it with the spectral power-law 
slope. The average index of flux power-law decay for this set is $\overline{\alpha} = 1.2 \pm 0.4$.
Together with that the celestial background is approximately $10^{-5.5}$ photons/s above 100 MeV, in a $5\deg$ region 
around the afterglow (the LAT PSF at 100 MeV), the above selection criterion implies that only afterglows brighter
than about $10^{-4}$ photons/s at peak or at first observation were retained.
The price to pay for that severe selection is the exclusion of 80 percent of afterglows seen by LAT, which were
too dim or decayed too fast. 

 Consequently, our selection criterion excludes dimmer afterglows with a fast decay and induces a brightness--decay correlation. 
 Still, as shown in Figure 4, six out of the seven brightest afterglows are fast-decayers, displaying a power-law decay 
index $\alpha > 1.2$ (the exception being 130427), thus the brightness-decay correlation is real.
A similar correlation has been observed for optical afterglows (Panaitescu \& Vestrand 2008) and has been attributed 
to how energy is deposited in the external-shock: short-lived ejecta outflows arrive quickly at the forward-shock and 
power bright afterglows, long-lived outflows deposit their energy slowlier and produce afterglows that are dimmer 
initially and which decay slower. 

 Therefore, the bright/fast-decaying--dim/slow-decaying dichotomy of LAT afterglows could be one piece of evidence for 
a long-lived energy-injection in some afterglows, a process which was found to occur in a large fraction of Swift X-ray 
afterglows (Panaitescu \& Vestrand 2012).
 About one third of the LAT afterglows in our sample show a light-curve decay slower than expected for the synchrotron
forward-shock emission and an impulsive ejecta release (Figure 4), although often consistent with it, due to the large 
uncertainties of the flux decay indices and/or spectral slopes. That is a second evidence for the energy-injection in 
afterglows that an extended ejecta-release would yield.

 The average slope of LAT afterglow spectra at lower energies (0.1-1 GeV) is $\overline{\beta} = 1.0 \pm 0.3$, 
with five afterglows (130327B, 131231, 150523, 150627, 160509) clearly being harder than $\beta = 1$ 
(i.e. their output peaks at a photon energy above LAT's range). 
Given that the inverse-Compton spectrum below the upscattered synchrotron cooling break is harder by $\Delta \beta = 0.5$ 
than the synchrotron emission above the cooling break, we propose that the LAT emission of the hardest afterglows is
inverse-Compton. 
A better evidence for the latter spectral component is provided by six afterglows (Table 2: 090328, 100414, 110721, 110731, 
130427, 140619B) whose spectrum hardens above a dip energy of $0.3-3$ GeV, indicating that synchrotron is dominant below 
the dip and inverse-Compton above it. 
In all, we find evidence for inverse-Compton emission being detected in 11 out of the 24 LAT afterglows considered here.

 The early spectrum of one afterglow displays a softening, which could be due to either photon-photon 
absorption in the source or to a maximal energy that electrons acquire in afterglow shocks.\footnote{
 Other mechanisms for a spectral softening -- the onset of Klein-Nishina regime for inverse-Compton scatterings 
 and attenuation by the EBL -- should both yield breaks above the LAT range} 
The former process softens the power-law $F(E) \propto E^{-\beta}$ afterglow spectrum by $\Delta \beta = \beta$
at the break-energy given in equation (\ref{Ec1}), the latter by a comparable $\Delta \beta = 1$ at the break-energy
given in equation (\ref{Ec2}). For each mechanism, a measurement of the break-energy determines the afterglow Lorentz 
factor (equations \ref{Gtau} and \ref{Gs}), while single power-law spectra setting only a lower-limit on $\Gamma$.

\acknowledgments{This work made use of the LAT data and Science Tools available at the Fermi Science Support Center,
    {\sl http://fermi.gsfc.nasa.gov/ssc/data/ }  }

\begin{table*}
 \caption{Light-curves of Figures 2 and 3. Flux $1\sigma$ upper limits (not tabulated) correspond to 1.9 photons
   for each time-bin}
\vspace*{4mm}
\centerline{\scriptsize
\begin{tabular}{l|llll|llll|lllllllllllllll}
  \hline
  GRB     & time & Flux    & $\pm\Delta t $ & $\sigma(F)$    & time   &  Flux   & $\pm\Delta t $ & $\sigma(F)$ &
            time & Flux    & $\pm\Delta t $ & $\sigma(F)$ \\
          &  (s) & ($cm^{-2} s^{-1}$) &  (s) & ($cm^{-2} s^{-1}$) & (s) & ($cm^{-2} s^{-1}$) &  (s) & ($cm^{-2} s^{-1}$) 
          &  (s) & ($cm^{-2} s^{-1}$) &  (s) & ($cm^{-2} s^{-1}$) \\
  \hline
  080916C & 3    & 1.4e-4 &  1  & 3.2e-4 &   6  & 4.3e-3 &  2   & 5.4e-4 &  12  & 1.4e-3 &  4  & 2.2e-4 \\
          & 24   & 1.0e-3 &  8  & 1.3e-4 &  48  & 3.4e-4 & 16   & 5.4e-5 &  96  & 1.5e-4 & 32  & 2.6e-5 \\
          & 192  & 5.2e-5 &  64 & 1.1e-5 & 384  & 1.6e-5 & 128  & 4.1e-6 & 768  & 7.2e-6 & 256 & 1.9e-6 \\
          & 1536 & 3.5e-6 & 512 & 1.0e-6 & 3072 & 4.4e-6 & 1024 & 2.4e-6 & 12288& 3.1e-6 & 4096& 5.9e-7 \\
 \hline
  090217  & 6    & 2.6e-4 &  2  & 1.7e-4 & 12   & 1.5e-4 & 4    & 9.2e-5 & 24   & 7.8e-5 & 8   & 4.7e-5 \\
          & 48   & 2.0e-5 & 16  & 1.9e-5 & 96   & 1.0e-5 & 32   & 1.0e-5 & 192  & 5.8e-6 & 64  & 5.7e-6 \\
          & 384  & 5.4e-6 & 128 & 4.3e-6 & 768  & 2.9e-6 & 256  & 6.7e-6 & 6144 & 1.7e-6 & 2048& 1.6e-6 \\
  \hline
  090328  & 12   & 1.5e-4 & 4   & 2.0e-4 & 24   & 1.1e-4 & 8    & 1.1e-4 & 48   & 3.3e-5 & 16  & 4.4e-5 \\
          & 96   & 4.9e-5 & 32  & 2.4e-5 & 192  & 2.0e-5 & 64   & 5.8e-6 & 384  & 1.9e-5 & 128 & 3.7e-6 \\
          & 768  & 8.4e-6 & 256 & 1.8e-6 & 1536 & 4.3e-6 & 512  & 8.8e-7 & 3072 & 3.5e-6 & 1024& 1.0e-6 \\
  \hline
  090510  & 0.75 & 3.2e-4 &0.25 & 7.3e-4 & 1.5  & 1.4e-2 & 0.5  & 1.5e-3 & 3    & 7.3e-3 & 1   & 7.6e-4 \\
          & 6    & 9.8e-4 & 2   & 2.0e-4 & 12   & 3.1e-4 & 4    & 7.8e-5 & 24   & 9.8e-5 & 8   & 4.2e-5 \\
          & 48   & 5.3e-5 & 16  & 1.6e-5 & 96   & 1.4e-5 & 32   & 8.5e-6 & 192  & 1.1e-5 & 64  & 4.8e-6 \\
          & 384  & 3.5e-6 & 128 & 2.1e-6 & 768  & 2.3e-6 & 256  & 1.2e-6 & 1536 & 2.7e-6 & 512 & 6.5e-7 \\
  \hline
  090902B & 0.75 & 5.1e-4 & 0.25& 1.2e-3 & 3    & 1.0e-3 & 1    & 5.0e-4 & 6    & 1.9e-3 & 2   & 3.5e-4 \\
          & 12   & 4.7e-3 & 4   & 3.9e-4 & 24   & 1.5e-3 & 8    & 1.6e-4 & 48   & 4.2e-4 & 16  & 5.5e-5 \\
          & 96   & 1.4e-4 & 32  & 2.0e-5 & 192  & 5.1e-5 & 64   & 7.8e-6 & 384  & 2.5e-5 & 128 & 3.9e-6 \\
          & 768  & 1.1e-5 & 256 & 1.8e-6 & 3072 & 5.3e-6 & 1024 & 1.2e-6 & 6144 & 3.9e-6 & 2048& 4.4e-7 \\
  \hline
  090926  & 3    & 2.5e-4 & 1   & 3.3e-4 & 6    & 3.3e-3 & 2    & 4.5e-4 & 12   & 5.3e-3 & 4   & 4.0e-4 \\
          & 24   & 1.1e-3 & 8   & 1.3e-4 & 48   & 3.3e-4 & 16   & 5.1e-5 & 96   & 1.1e-4 & 32  & 2.1e-5 \\
          & 192  & 4.8e-5 & 64  & 1.1e-5 & 3072 & 4.8e-6 & 1024 & 9.0e-7 & 6144 & 4.6e-6 & 2048& 7.2e-7 \\
  \hline
  091003  & 1.5  & 1.8e-4 & 0.5 & 4.1e-4 & 6    & 1.3e-4 & 2    & 1.3e-4 & 12   & 4.4e-5 & 4   & 5.9e-5 \\
          & 24   & 2.2e-5 & 8   & 2.9e-5 & 48   & 2.2e-5 & 16   & 1.8e-5 & 96   & 1.4e-5 & 32  & 9.5e-6 \\
          & 192  & 7.1e-6 & 64  & 4.8e-6 & 384  & 8.3e-6 & 128  & 2.4e-6 & 768  & 2.7e-6 & 256 & 1.3e-6 \\
          & 1536 & 2.5e-6 & 512 & 1.33-6 & 6144 & 1.6e-6 & 2048 & 3.0e-7 &12288 & 1.8e-6 & 4096& 2.8e-7 \\
  \hline
  100414  & 15   & 7.6e-5 & 5   & 1.8e-4 & 30   & 2.9e-4 & 10   & 1.4e-4 & 60   & 1.5e-4 & 20  & 4.3e-5 \\
          & 120  & 2.9e-5 & 40  & 1.3e-5 & 240  & 1.7e-5 & 80   & 4.2e-6 & 480  & 5.7e-6 & 160 & 1.6e-6 \\
          & 960  & 1.9e-6 & 320 & 9.2e-7 & 1920 & 3.0e-6 & 640  & 6.3e-7 & 3840 & 3.3e-6 & 1280& 6.8e-7 \\
  \hline
  110721  & 0.75 & 3.1e-3 & 0.25& 1.6e-3 & 1.5  & 2.0e-3 & 0.5  & 8.4e-4 & 3    & 4.9e-4 & 1   & 3.3e-4 \\
          & 6    & 2.4e-4 & 2   & 1.6e-4 & 12   & 9.7e-5 & 4    & 7.7e-5 & 24   & 4.8e-5 & 8   & 3.8e-5 \\
          & 48   & 1.7e-5 & 16  & 1.7e-5 & 96   & 1.0e-5 & 32   & 8.0e-6 & 192  & 7.1e-6 & 64  & 4.2e-6 \\
          & 384  & 5.4e-6 & 128 & 2.5e-6 & 768  & 2.7e-6 & 256  & 1.3e-6 & 3072 & 2.1e-6 & 1024& 4.1e-7 \\
  \hline
  110731  & 4    & 8.5e-4 & 1   & 3.6e-4 & 7.5  & 1.2e-3 & 2.5  & 2.0e-4 & 15   & 1.4e-4 & 5   & 6.8e-5 \\
          & 30   & 5.1e-5 & 10  & 3.1e-5 & 60   & 2.1e-5 & 20   & 1.4e-5 & 120  & 6.4e-6 & 40  & 6.2e-6 \\
          & 240  & 3.2e-6 & 80  & 3.2e-6 & 480  & 4.3e-6 & 160  & 2.1e-6 & 960  & 3.3e-6 & 320 & 9.4e-7 \\
  \hline
  130327B & 7.5  & 5.2e-5 & 2.5 & 1.2e-4 & 30   & 1.9e-4 & 10   & 4.9e-5 & 60   & 7.9e-5 & 20  & 2.2e-5 \\
          & 120  & 3.5e-5 & 40  & 9.6e-6 & 240  & 5.5e-6 & 80   & 3.8e-6 & 480  & 4.9e-6 & 160 & 2.2e-6 \\
          & 1920 & 3.6e-6 & 640 & 8.3e-6 & 3840 & 5.5e-6 & 1280 & 2.3e-6 & 7680 & 1.4e-6 & 2560& 5.9e-7 \\
  \hline
  130427  & 5.5  & 5.8e-4 & 4.5 & 1.4e-4 & 15   & 1.8e-3 & 5    & 2.4e-4 & 30   & 1.2e-3 & 10  & 1.4e-4 \\
          & 60   & 4.8e-4 & 20  & 6.0e-5 & 120  & 2.9e-4 & 40   & 3.0e-5 & 240  & 1.7e-4 & 80  & 1.5e-5 \\
          & 480  & 8.7e-5 & 160 & 8.2e-6 & 3840 & 5.9e-6 & 1280 & 7.9e-7 & 7680 & 3.2e-6 & 2560& 5.4e-7 \\
  \hline
  130821  & 48   & 3.2e-5 & 16  & 2.2e-5 & 96   & 2.5e-5 & 32   & 1.2e-5 & 192  & 1.7e-5 & 64  & 4.9e-6 \\
          & 384  & 1.6e-5 & 128 & 3.2e-6 & 768  & 5.2e-6 & 256  & 1.3e-6 & 1536 & 3.6e-6 & 512 & 8.0e-7 \\
          & 3072 & 2.8e-6 & 1024& 2.2e-6 & 6144 & 2.2e-6 & 2048 & 3.6e-7 & 12288& 1.1e-6 & 4096& 2.9e-7 \\
  \hline
  131108  & 1.5  & 2.8e-3 & 0.5 & 6.7e-4 & 3.5  & 1.2e-3 & 1.5  & 2.5e-4 & 7.5  & 2.1e-3 & 2.5 & 2.6e-4 \\
          & 15   & 4.8e-4 & 5   & 8.8e-5 & 30   & 1.5e-4 & 10   & 3.4e-5 & 60   & 3.6e-5 & 20  & 1.6e-5 \\
          & 120  & 7.8e-6 & 40  & 6.2e-6 & 240  & 4.6e-6 & 80   & 3.1e-6 & 480  & 5.6e-6 & 160 & 1.6e-6 \\
  \hline
  131231  & 24   & 3.0e-5 & 8   & 4.1e-5 & 48   & 5.3e-5 & 16   & 2.9e-5 & 96   & 2.5e-5 & 32  & 1.4e-5 \\
          & 192  & 1.9e-5 & 64  & 5.3e-6 & 384  & 4.5e-6 & 128  & 2.7e-6 & 768  & 3.5e-6 & 256 & 1.6e-6 \\
  \hline
  140206  & 12   & 2.3e-4 & 4   & 1.2e-4 & 24   & 1.5e-4 & 8    & 6.7e-5 & 48   & 1.6e-5 & 16  & 2.2e-5 \\
          & 96   & 2.6e-5 & 32  & 1.4e-5 & 192  & 4.6e-6 & 64   & 4.5e-6 & 384  & 8.8e-6 & 128 & 2.4e-6 \\
          & 768  & 4.8e-6 & 256 & 1.3e-6 & 1536 & 3.8e-6 & 512  & 8.3e-7 & 3072 & 4.6e-6 & 1024& 9.8e-7 \\
  \hline
  140523  & 1.5  & 3.4e-4 & 0.5 & 7.9e-4 & 3    & 3.4e-4 & 1    & 4.5e-4 & 6    & 8.7e-5 & 2   & 2.0e-4 \\
          & 12   & 2.7e-4 & 4   & 1.6e-4 & 24   & 2.4e-4 & 8    & 7.3e-5 & 48   & 9.1e-5 & 16  & 4.5e-5 \\
          & 96   & 2.4e-5 & 32  & 1.9e-5 & 192  & 3.3e-5 & 64   & 1.5e-5 & 384  & 3.0e-5 & 128 & 1.3e-5 \\
  \hline
  140619B & 0.17 & 2.8e-3 &0.075& 3.7e-3 & 0.37 & 5.8e-3 & 0.12 & 3.1e-3 & 0.75 & 2.5e-3 & 0.25& 1.5e-3 \\
          & 1.5  & 1.0e-3 & 0.5 & 7.1e-4 & 3    & 2.1e-4 & 1    & 2.8e-4 & 12   & 2.6e-5 & 4   & 6.0e-5 \\
          & 24   & 1.3e-5 & 8   & 3.0e-5 & 192  & 1.5e-6 & 64   & 3.6e-6 & 384  & 7.6e-7 & 128 & 1.8e-6 \\
  \hline
  141028  & 0.75 & 4.0e-4 & 0.25& 9.3e-4 & 3    & 3.0e-4 & 1    & 2.9e-4 & 6    & 1.5e-4 & 2   & 1.5e-4 \\
          & 12   & 1.0e-4 & 4   & 7.9e-5 & 24   & 2.5e-5 & 8    & 3.3e-5 & 48   & 2.4e-5 & 16  & 1.9e-5 \\
          & 96   & 1.2e-5 & 32  & 9.3e-6 & 192  & 1.3e-5 & 64   & 7.2e-6 & 384  & 8.2e-6 & 128 & 5.6e-6 \\
  \hline
  150523  & 3    & 9.5e-5 & 1   & 2.2e-4 & 6    & 9.5e-5 & 2    & 1.3e-4 & 12   & 1.9e-4 & 4   & 9.4e-5 \\
          & 24   & 7.1e-5 & 8   & 4.3e-5 & 48   & 5.4e-5 & 16   & 2.5e-5 & 96   & 2.4e-5 & 32  & 1.2e-5 \\
          & 192  & 1.4e-5 & 64  & 4.3e-6 & 384  & 5.7e-6 & 128  & 2.6e-6 & 768  & 3.6e-6 & 256 & 1.1e-6 \\
  \hline
  150627  & 24   & 1.2e-4 & 8   & 2.9e-4 & 48   & 5.2e-5 & 16   & 1.2e-4 & 96   & 3.9e-5 & 32  & 3.8e-5 \\
          & 192  & 2.9e-5 & 64  & 7.4e-6 & 384  & 8.0e-6 & 128  & 2.3e-6 & 768  & 5.3e-6 & 256 & 1.3e-6 \\
          & 1536 & 4.4e-6 & 512 & 8.9e-7 & 3072 & 2.4e-6 & 1024 & 5.0e-7 & 6144 & 3.2e-6 & 2048& 4.2e-7 \\
  \hline
  150902  & 6    & 4.5e-4 & 2   & 2.2e-4 & 12   & 2.5e-4 & 4    & 1.2e-4 & 24   & 1.1e-4 & 8   & 5.5e-5 \\
          & 48   & 2.8e-5 & 16  & 2.2e-5 & 96   & 2.9e-5 & 32   & 1.3e-5 & 192  & 1.3e-5 & 64  & 6.1e-6 \\
          & 384  & 1.2e-5 & 128 & 2.9e-6 & 768  & 4.4e-6 & 256  & 1.3e-6 & 1536 & 3.7e-6 & 512 & 7.8e-7 \\
  \hline
  160325  & 6    & 4.4e-5 & 2   & 1.0e-4 & 48   & 1.1e-5 & 16   & 1.5e-5 & 96   & 2.9e-5 & 32  & 1.2e-5 \\
          & 192  & 8.6e-6 & 64  & 5.1e-6 & 384  & 4.6e-6 & 128  & 2.7e-6 & 768  & 3.2e-6 & 256 & 1.6e-6 \\
  \hline
  160509  & 12   & 5.9e-4 & 4   & 1.3e-4 & 24   & 8.7e-4 & 8    & 1.1e-4 & 48   & 2.6e-5 & 16  & 2.1e-5 \\
          & 96   & 2.2e-5 & 32  & 1.2e-5 & 192  & 1.6e-5 & 64   & 4.9e-6 & 384  & 5.3e-6 & 128 & 2.8e-6 \\
          & 768  & 6.0e-6 & 256 & 1.5e-6 & 1536 & 2.5e-6 & 512  & 6.5e-7 & 3072 & 1.5e-6 & 1024& 1.5e-6 \\
  \hline
\end{tabular}
}
\end{table*}

\clearpage

\begin{table*}
 \caption{Spectra of Figures 2 and 3 ($1\sigma$ upper limits, corresponding to 1.9 photons, are not tabulated).
     Fluxes are in $erg/cm^2s$; $1\sigma$ upper uncertainty $\sigma_+$ is given in parenthesis. 
     For fluxes above several photons per energy channel per time-bin, upper and lower uncertainties are 
     about equal ($\sigma_+ \simeq \sigma_-$) but, for fluxes of a few photons, $\sigma_- < \sigma_+$. }
\vspace*{4mm}
\centerline{\small
\begin{tabular}{llllllllllllllllllllllll}
  \hline
  GRB     & time  & F(0.14k) & F(0.28k) & F(0.56k) & F(1.1k)  &  F(2.2k) & F(4.5k)  & F(8.9k)  & F(18k)   & F(35k)   & F(71k) \\
          & (s)   & ($\sigma_+$) & ($\sigma_+$) & ($\sigma_+$) & ($\sigma_+$) & ($\sigma_+$) & ($\sigma_+$) & ($\sigma_+$)
          & ($\sigma_+$) & ($\sigma_+$) & ($\sigma_+$) \\
 \hline  
  080916C &10-100 &  1.2e-7  &  1.0e-7  &  1.4e-7  &  1.2e-7  &  7.7e-8  &  3.8e-8  &  1.5e-7  &          &  3.1e-7  &        \\
          &       & (1.8e-8) & (1.9e-8) & (2.9e-8) & (3.3e-8) & (6.1e-8) & (8.9e-8) & (2.0e-7) &          & (7.1e-7) &        \\
  \hline
  090217  & 0-100 &  1.2e-8  &  1.5e-8  &  1.2e-8  &  6.6e-9  &          &          &          &          &          &        \\
          &       & (4.6e-9) & (6.1e-9) & (1.1e-8) & (1.5e-8) &          &          &          &          &          &        \\
  \hline
  090328  &10-1000&  4.7e-9  &  4.2e-9  &  3.1e-9  &  2.0e-9  &  5.2e-9  &  1.3e-8  &  5.2e-9  &          &          &        \\
          &       & (9.0e-10)& (1.0e-9) & (1.1e-9) & (2.0e-9) & (4.1e-9) & (8.9e-9) & (1.2e-8) &          &          &        \\
  \hline
  090510  & 0-10  &  6.3e-7  &  8.8e-7  &  7.6e-7  &  8.6e-7  &  1.3e-6  &  6.4e-7  &  4.2e-7  &          &  1.7e-6  &        \\
          &       & (9.4e-8) & (1.3e-7) & (1.6e-7) & (2.1e-7) & (3.7e-7) & (6.3e-7) & (9.8e-7) &          & (3.9e-6) &        \\
  \hline
  090902B & 10-100&  9.3e-8  &  2.1e-7  &  1.8e-7  &  1.9e-7  &  3.5e-7  &  2.4e-7  &  1.4e-7  &  5.5e-7  &  2.7e-7  &        \\
          &       & (1.5e-8) & (2.6e-8) & (3.1e-8) & (4.1e-8) & (7.7e-8) & (1.3e-7) & (1.8e-7) & (4.4e-7) & (6.4e-7) &        \\
  090902B &100-1000& 7.4e-9  &  6.5e-9  &  5.1e-9  &  5.7e-9  &  5.7e-9  &  1.4e-8  &          &  1.1e-8  &          &        \\
          &       & (1.2e-9) & (1.3e-9) & (1.5e-9) & (2.8e-9) & (4.5e-9) & (9.6e-9) &          & (2.6e-8) &          &        \\
  \hline
  090926  & 10-100&  1.9e-7  &  2.6e-7  &  1.9e-7  &  1.9e-7  &  2.5e-7  &  1.6e-7  &  8.2e-8  &  1.6e-7  &          &        \\
          &       & (2.3e-8) & (3.2e-8) & (3.4e-8) & (4.4e-8) & (7.1e-8) & (1.3e-7) & (1.9e-7) & (3.8e-7) &          &        \\
  \hline
  090103  & 0-100 &  8.5e-9  &  6.1e-9  &  6.3e-9  &  5.3e-9  &  2.1e-8  &          &          &          &          &        \\
          &       & (3.5e-9) & (6.0e-9) & (8.4e-9) & (1.2e-8) & (2.8e-8) &          &          &          &          &        \\
  \hline
  100414  &10-1000&  2.2e-9  &  2.6e-9  &  1.7e-9  &  2.9e-9  &  1.1e-9  &  4.5e-9  &          &          &  3.6e-8  &        \\
          &       & (5.9e-10)& (7.5e-10)& (1.1e-9) & (1.9e-9) & (2.6e-9) & (6.0e-9) &          &          & (4.8e-8) &        \\
  \hline
  110721  &  0-10 &  2.5e-7  &  7.7e-8  &  4.0e-8  &  1.3e-7  &          &          & 5.3e-7   &          &          &        \\
          &       & (6.7e-8) & (7.5e-8) & (9.3e-8) & (1.8e-7) &          &          & (1.2e-6) &          &          &        \\
  \hline
  110731  & 0-100 &  4.7e-8  &  2.9e-8  &  1.8e-8  &  3.6e-8  & 4.1e-8   &          &          &          &          &        \\
          &       & (8.0e-9) & (7.5e-9) & (7.4e-9) & (1.4e-8) & (3.2e-8) &          &          &          &          &        \\
  \hline
  130327B & 0-150 &  9.5e-9  &  1.5e-8  &  1.3e-8  &  2.2e-8  &  3.6e-8  &  3.6e-8  &  3.6e-8  &          &          &        \\
          &       & (3.4e-9) & (5.1e-9) & (9.0e-9) & (1.5e-8) & (2.8e-8) & (4.8e-8) & (8.3e-8) &          &          &        \\
  \hline
  130427  & 1-20  &  3.5e-7  &  2.5e-7  &  2.7e-7  &  1.5e-7  &  2.0e-7  &          &  7.9e-7  &          &          & 3.2e-6 \\
          &       & (6.8e-8) & (6.9e-8) & (1.2e-7) & (1.5e-7) & (2.6e-7) &          & (1.0e-6) &          &          & (7.3e-6) \\
  130427  & 20-100&  1.1e-7  &  1.5e-7  &  1.6e-7  &  2.2e-7  &  2.2e-7  &  2.4e-7  &  4.8e-7  &  1.9e-7  &  1.2e-6  &        \\
          &       & (1.9e-8) & (2.6e-8) & (3.4e-8) & (5.1e-8) & (1.0e-7) & (1.6e-7) & (3.3e-7) & (4.4e-7) & (1.1e-6) &        \\
  130427  &100-600&  4.3e-8  &  4.3e-8  &  3.7e-8  &  3.7e-8  &  1.7e-8  &  4.0e-8  &  4.0e-8  &  5.4e-8  &  5.3e-8  & 2.1e-7 \\
          &       & (4.4e-9) & (5.2e-9) & (6.1e-9) & (7.9e-9) & (1.1e-8) & (2.4e-8) & (3.9e-8) & (7.1e-8) & (1.2e-7) & (2.8e-7) \\
  \hline
  130821  &30-1000&  5.0e-9  &  4.8e-9  &  5.3e-9  &  1.8e-9  &  1.8e-9  &  3.5e-9  &          &          &          &        \\
          &       & (1.1e-9) & (1.3e-9) & (1.7e-9) & (2.4e-9) & (4.1e-9) & (8.2e-9) &          &          &          &        \\
  \hline
  131108  & 10-100&  3.9e-8  &  2.8e-8  &  1.1e-8  &  1.2e-8  &          &          &          &          &          &        \\
          &       & (7.9e-9) & (8.0e-9) & (1.1e-8) & (1.6e-8) &          &          &          &          &          &        \\
  \hline
  131231  &10-1000&  5.3e-9  &  5.9e-9  &  7.9e-9  &  4.5e-9  &  8.9e-9  &  1.8e-8  &  1.8e-8  &          &  7.0e-8  &        \\
          &       & (1.8e-9) & (2.2e-9) & (3.2e-9) & (5.9e-9) & (1.2e-8) & (2.4e-8) & (4.1e-8) &          & (1.6e-7) &        \\
  \hline
  140206  & 10-500&  4.2e-9  &  3.0e-9  &  3.9e-9  &          &  2.6e-9  &  5.2e-9  &          &          &          &        \\
          &       & (1.2e-9) & (1.2e-9) & (2.6e-9) &          & (6.0e-9) & (1.2e-8) &          &          &          &        \\
  \hline
  140523  & 10-500&  2.2e-9  &  1.5e-8  &  1.2e-8  &  1.2e-8  &  2.4e-8  &          &  3.2e-8  &          &          &        \\
          &       & (2.9e-9) & (4.9e-9) & (8.2e-9) & (1.2e-8) & (2.4e-8) &          & (7.5e-8) &          &          &        \\
  \hline
  140619B &  0-20 &  5.9e-8  &  5.6e-8  &  6.6e-8  &  3.7e-8  &          &          &          &  5.9e-7  &          &        \\
          &       & (2.4e-8) & (4.5e-8) & (6.4e-8) & (8.6e-8) &          &          &          & (1.4e-6) &          &        \\
  \hline
  141028  & 0-500 &  3.2e-9  &  4.1e-9  &  3.6e-9  &  3.1e-9  &          &  6.1e-9  &          &          &          &        \\
          &       & (1.1e-9) & (1.5e-9) & (2.9e-9) & (4.1e-9) &          & (1.4e-8) &          &          &          &        \\
  \hline
  150523  & 0-1000&  2.3e-9  &  2.4e-9  &  3.4e-9  &  2.8e-9  &  4.5e-9  &  4.5e-9  &  4.5e-9  &          &          &        \\
          &       & (5.8e-10)& (7.1e-10)& (1.1e-9) & (1.9e-9) & (3.6e-9) & (6.0e-9) & (1.0e-8) &          &          &        \\
  \hline
  150627  &10-1000&  8.8e-10 &  1.8e-9  &  1.2e-9  &  2.6e-9  &  6.6e-9  &  1.3e-8  &  5.2e-9  &          &          &        \\
          &       & (6.0e-10)& (6.6e-10)& (1.1e-9) & (2.1e-9) & (4.5e-9) & (8.9e-9) & (1.2e-8) &          &          &        \\
  \hline
  150902  & 5-500 &  6.6e-9  &  6.9e-9  &  3.8e-9  &  2.6e-9  &  5.2e-9  &  5.2e-9  &  1.0e-8  &          &          &        \\
          &       & (1.5e-9) & (1.9e-9) & (2.6e-9) & (3.4e-9) & (6.9e-9) & (1.2e-8) & (2.4e-8) &          &          &        \\
  \hline
  160325  &100-1000& 1.2e-9  &  8.7e-10 &  1.4e-9  &  4.6e-10 &  1.8e-9  &          &          &          &          &        \\
          &       & (5.2e-10)& (5.9e-10)& (9.2e-10)& (1.1e-9) & (2.4e-9) &          &          &          &          &        \\
  \hline
  160509  &33-1000&  2.1e-9  &  3.6e-9  &  2.1e-9  &  5.4e-9  &  4.7e-9  &  4.7e-9  &  4.7e-9  &          &  1.9e-8  & 3.8e-8 \\
          &       & (5.7e-10)& (9.1e-10)& (1.3e-9) & (2.5e-9) & (3.8e-9) & (6.3e-9) & (1.1e-8) &          & (4.4e-8) & (8.7e-8) \\
  \hline
\end{tabular}
}
\end{table*}


\begin{references}
\reference{} Abdo A. et al, 2009a, Sci 323, 1688
\reference{} Abdo A. et al, 2009b, ApJ 706, L138
\reference{} Ackermann M. et al, 2010a, ApJ 716, 1178
\reference{} Ackermann M. et al, 2010b, ApJ 717, L127
\reference{} Ackermann M. et al, 2011, ApJ 729, 114
\reference{} Ackermann M. et al, 2013a, ApJS 209, 11
\reference{} Ackermann M. et al, 2013b, ApJ 763, 71
\reference{} Ackermann M. et al, 2014, Sci 343, 42
\reference{} Atwood, W. et al, 2009, ApJ 697, 1071
\reference{} Axelsson M. et al 2012, ApJ 757, L31
\reference{} Fan Y-Z et al, 2013, ApJ 776, 95
\reference{} Ghirlanda G., Ghisellini G., Nava L., 2010, A\&A 510, L7
\reference{} Giuliani A. et al, 2010, ApJ 708, L84
\reference{} Giuliani A. et al, 2014, A\&AL, submitted (lanl.arxiv.org/abs/1407.0238)
\reference{} Kumar P., Barniol Duran R., 2009, MNRAS 409, 226 
\reference{} Kumar P., Zhang B., 2015, PhR 561, 1
\reference{} Liang, Y-F et al, 2014, ApJ 781, L74 (2014)
\reference{} Liu B. et al, 2014, ApJ 787, L6
\reference{} M\'esz\'aros P., Rees M., 1997, ApJ 476, 232
\reference{} Nousek J. et al, 2006, ApJ, 642, 389
\reference{} Panaitescu A., 2007, MNRAS 379, 331 
\reference{} Panaitescu A., Vestrand T., 2008, MNRAS 387, 497
\reference{} Panaitescu A., Vestrand T., 2011, MNRAS 414, 3537
\reference{} Panaitescu A., Vestrand T., 2012, MNRAS 425, 1669
\reference{} Panaitescu A., Vestrand T., Wozniak P., 2013, MNRAS 433, 759
\reference{} Panaitescu A., 2015, ApJ 806, 64
\reference{} Rees M., M\'esz\'aros P., 1998, ApJ 496, L1
\reference{} Swenson C. et al., 2010, ApJ 718, L14
\reference{} Tam P.-H. et al, 2013, ApJ 772, L4
\end{references}
\end{document}